\documentclass[12pt]{article} 

\begin{document}

\baselineskip=12pt plus 1pt minus 1pt 

\begin{center}

{\large \bf Rotationally invariant Hamiltonians for nuclear spectra 
\\ based on quantum algebras}  

\bigskip\bigskip\bigskip 

{Dennis Bonatsos$^{\#}$\footnote{e-mail: bonat@inp.demokritos.gr}, 
B. A. Kotsos$^*$\footnote{e-mail: bkotsos@teilam.gr}, 
P. P. Raychev$^\dagger$\footnote{e-mail: raychev@phys.uni-sofia.bg, 
raychev@inrne.bas.bg}, 
P. A. Terziev$^\dagger$\footnote{e-mail: terziev@inrne.bas.bg} }

\bigskip 
{$^{\#}$ Institute of Nuclear Physics, N.C.S.R. ``Demokritos'', \\ GR-15310 
Aghia Paraskevi, Attiki, Greece}

\medskip
{$^*$ Department of Electronics, Technological Education Institute, 
\\ GR-35100 Lamia, Greece} 

\medskip
{$^\dagger$ Institute for Nuclear Research and Nuclear Energy, Bulgarian
Academy of Sciences, \\ 72 Tzarigrad Road, BG-1784 Sofia, Bulgaria}

\vskip 0.5truein 

\centerline{\bf Abstract} 

\end{center} 

The rotational invariance under the usual physical angular momentum of the 
su$_q$(2) Hamiltonian for the description of rotational nuclear spectra 
is explicitly proved and a connection of this Hamiltonian to the formalisms 
of Amal'sky and Harris is provided. In addition, a new Hamiltonian for 
rotational spectra is introduced, based on the construction of irreducible 
tensor operators (ITO) under su$_q$(2) and use of $q$-deformed tensor products 
and $q$-deformed Clebsch--Gordan coefficients. The rotational invariance 
of this su$_q$(2) ITO Hamiltonian under the usual physical angular momentum 
is explicitly proved, a simple closed expression for its energy spectrum 
(the ``hyperbolic tangent formula'') is introduced, and its connection 
to the Harris formalism is established. Numerical tests in a series of 
Th isotopes are provided.  

\bigskip\bigskip 



\section{Introduction}

Quantum algebras \cite{Chari,Bieden,Klimyk} 
have started finding applications in the description 
of symmetries of physical systems over the last years \cite{PPNP}. 
In one of the earliest 
attempts, a Hamiltonian proportional to the second order Casimir operator 
of su$_q$(2) has been used for the description of rotational nuclear 
spectra \cite{RRS} and its relation to the Variable Moment of Inertia Model 
\cite{PLB251} has been clarified. 

However, several open problems remained:

a) Is the su$_q$(2) Hamiltonian invariant under the usual su(2) Lie 
algebra, i.e.  under usual angular momentum, or it breaks spherical 
symmetry and/or the isotropy of space? 

b) How does the physical angular momentum appear in the framework 
of su$_q$(2)? Is there any relation between the generators of su$_q$(2) 
and the usual physical angular momentum operators? 

c) How can one add angular momenta in the su$_q$(2) framework? In other words,
how does angular momentum conservation work in the su$_q$(2) framework? 

Answers to these questions are provided in the present paper, along with 
connections of the su$_q$(2) model to other formalisms. 

After a brief introduction to the su$_q$(2) formalism in Section 2, we 
prove explicitly in Section 3 that the su$_q$(2) Hamiltonian does 
commute with the generators of su(2), i.e. with the generators of usual 
physical angular momentum. Therefore the su$_q$(2) Hamiltonian does not 
violate the isotropy of space and does not destroy spherical symmetry. 
The generators of su$_q$(2) are expressed in terms of the generators 
of su(2). In addition, it turns out that the angular momentum quantum numbers 
appearing in the description of the irreducible representations (irreps)
of su$_q$(2) are exactly the same as the ones appearing in the irreps 
of su(2), establishing an one-to-one correspondence between the two sets of
irreps (in the generic case in which the deformation parameter $q$ is not 
a root of unity). 

Taking advantage of the results of Section 3, we write in Section 4 the 
eigenvalues of the su$_q$(2) Hamiltonian as an exact power series in 
$l(l+1)$ (where $l$ is the usual physical angular momentum). An approximation 
to this expansion, studied in Section 5, leads to a closed energy formula for 
rotational spectra introduced by Amal'sky \cite{Amal}. The study of analytic 
expressions for the moment of inertia and the rotational frequency
based on the closed formula of Section 5 leads in Section 6 to a connection 
between the present approach and the Harris formalism \cite{Harris}. 

We then turn in Section 7 into the study of irreducible tensor operators 
under su$_q$(2) \cite{STK593,STK690}, 
constructing the irreducible tensor operator of rank one 
corresponding to the su$_q$(2) generators. We also define tensor products 
in the su$_q$(2) framework and construct the scalar square of the angular 
momentum operator, a task requiring the use of $q$-deformed Clebsch--Gordan 
coefficients \cite{STK593}. In addition to exhibiting explicitly how addition 
of angular momenta works in the su$_q$(2) framework, this exercise leads 
to a Hamiltonian built out of the components of the above mentioned 
irreducible tensor operator (ITO), which can also be applied to the 
description of rotational spectra. We are going to refer to this Hamiltonian 
as the {\sl  su$_q$(2) ITO Hamiltonian}. 

The fact that the su$_q$(2) ITO Hamiltonian does commute with the generators 
of the usual su(2) algebra is shown explicitly in Section 8. Based on the 
results of Section 8, we express in Section 9 the eigenvalues of the su$_q$(2)
ITO Hamiltonian as an exact power series in $l(l+1)$, where $l$ is the 
usual physical angular momentum. An approximation to this series, 
studied in Section 10, leads to a simple closed formula for the spectrum
(the ``hyperbolic tangent formula''),
which is used in Section 11 in order to obtain analytic expressions for the 
moment of inertia and the rotational frequency, leading to a connection 
of the present results to the Harris formalism \cite{Harris}. 

Finally in Section 12 all the exact and closed approximate energy formulae 
obtained above are compared to the experimental spectra of a series 
of Th isotopes, as well as to the results provided by the usual 
rotational expansion and by the Holmberg--Lipas formula \cite{Lipas}, which 
is probably the best two-parameter formula for the description 
of rotational nuclear spectra \cite{Casten}. A discusion of the present 
results and plans for future work are given in Section 13. 

\section{The quantum algebra su$_q$(2)} 

The quantum algebra su$_q$(2) \cite{KulRes,Skly,Jimbo} 
is a $q$-deformation of the Lie algebra su(2).
It is generated by the operators $L_{+}$, $L_{-}$, $L_{0}$, obeying the 
commutation relations (see \cite{PPNP} and references therein) 
\begin{equation} \label{eq:1.1}
[L_0,L_{\pm}]=\pm L_{\pm},
\end{equation} %
\begin{equation} \label{eq:1.2}
[L_{+},L_{-}]=[2L_0]=\frac{q^{2L_0}-q^{-2L_0}}{q-q^{-1}},
\end{equation} %
where $q$-numbers and $q$-operators are defined by 
\begin{equation} \label{eq:1.3}
[x] = {q^x -q^{-x} \over q-q^{-1} } .
\end{equation} %
There are two distinct cases for the domain of the deformation parameter: 

a) $q=e^\tau$, $\tau\in{\bf R}$, in which
\begin{equation} \label{eq:1.4}
[x]= {\sinh{\tau x}\over \sinh{\tau} }, 
\end{equation} %

b) $q=e^{i\tau}$, $\tau\in{\bf R}$, in which 
\begin{equation} \label{eq:1.5}
[x]={\sin{\tau x} \over \sin{\tau}}.
\end{equation} %

In both cases one has 
\begin{equation} \label{eq:1.6}
[x]\rightarrow x \quad {\rm as} \quad q\rightarrow 1. 
\end{equation}

If the deformation parameter $q$ is not a root of unity
[$q$ is a root of unity in case b) if one has $q^n = 1$, $n\in{\bf N}$]
the finite-dimensional irreducible representation $D^\ell_{(q)}$ of su$_q$(2) 
is determined by the highest weight vector $|\ell,\ell\rangle_q$ with 
\begin{equation} \label{eq:1.7}
L_{+}|\ell,\ell\rangle_q=0,
\end{equation}  %
and the basis states $|\ell,m\rangle_q$ are expressed as 
\begin{equation} \label{eq:1.8}
|\ell,m\rangle_q=\sqrt{\frac{[\ell+m]!}{[2\ell]![\ell-m]!}}
\,(L_{-})^{\ell-m}\,|\ell,\ell\rangle_q, 
\end{equation} %
where $[n]!=[n][n-1]\ldots[1]$ is the notation for the $q$-factorial. Then the
explicit form of the irreducible representation (irrep) $D^\ell_{(q)}$ of the 
su$_q$(2) algebra is determined by the equations
\begin{equation} \label{eq:1.9}
L_{\pm}|\ell,m\rangle_q=\sqrt{[\ell\mp m][\ell\pm m+1]}\,|\ell,m\pm1\rangle_q, 
\end{equation} %
\begin{equation} \label{eq:1.10}
L_0|\ell,m\rangle_q=m\,|\ell,m\rangle_q, 
\end{equation} %
and the dimension of the corresponding representation is the same as in the
non-deformed case, i.e.  ${\rm dim} D^\ell_{(q)}=2\ell+1$ for 
$\ell=0,\frac{1}{2},1,\frac{3}{2},2\ldots$

The second-order Casimir operator of su$_q$(2) is 
$$ C_2^{(q)}=\frac{1}{2}(L_{+}L_{-}+L_{-}L_{+}+[2][L_{0}]^2) $$
\begin{equation} \label{eq:1.11}
=L_{-}L_{+}+[L_0][L_0+1]=L_{+}L_{-}+[L_0][L_0-1],
\end{equation} %
while its eigenvalues in the space of the irreducible 
representation $D^\ell_{(q)}$ are $[\ell][\ell+1]$
\begin{equation} \label{eq:1.12}
C_2^{(q)} | \ell,m\rangle_q = [\ell][\ell+1] |\ell,m\rangle_q.
\end{equation}

It has been suggested (see \cite{PPNP,RRS} and references therein) 
that rotational spectra of deformed nuclei and diatomic molecules can be 
described by a phenomenological Hamiltonian
based on the symmetry of the quantum algebra su$_q$(2)
\begin{equation} \label{eq:1.13}
H=\frac{\hbar^2}{2{\cal J}_0}\,C_{2}^{(q)}+E_0, 
\end{equation} %
where $C_{2}^{(q)}$ is the second order Casimir operator of Eq. 
(\ref{eq:1.11}), ${\cal J}_0$ is the moment of inertia for the non-deformed 
case $q\to1$, and $E_0$ is the bandhead energy for a given band.

The eigenvalues of the Hamiltonian of Eq. (\ref{eq:1.13}) 
in the basis of Eq. (\ref{eq:1.8}) are then
\begin{equation} \label{eq:1.14}
E_{\ell}^{(\tau)} = A [\ell] [\ell+1] +E_0,
\end{equation} %
where the definition 
\begin{equation} \label{eq:1.15}
A={\hbar^2 \over 2 {\cal J}_0}
\end{equation}  %
has been used for brevity. 

In the case with $q=e ^\tau,\ \tau\in{\bf R}$ the spectrum of the model 
Hamiltonian of Eq. (\ref{eq:1.13})  takes the form
\begin{equation} \label{eq:1.16}
E_\ell^{(\tau)}=A \,
\frac{\sinh(\ell\tau)\sinh((\ell+1)\tau)}{\sinh^2(\tau)}+E_0, 
\qquad q=e^{\tau} 
\end{equation} %
while in the case with $q=e^{i\tau}$, $\tau\in{\bf R}$ and 
$q^n\neq 1$, $n\in{\bf N}$ 
the spectrum of the model Hamiltonian of Eq. (\ref{eq:1.13})  
takes the form 
\begin{equation} \label{eq:1.17}
E_\ell^{(\tau)}=A \,
\frac{\sin(\ell\tau)\sin((\ell+1)\tau)}{\sin^2(\tau)}+E_0, 
\qquad q=e^{i\tau}. 
\end{equation} %

It is known (see \cite{PPNP,RRS} and references therein) that 
only the spectrum of Eq. (\ref{eq:1.17}) exhibits behavior that is in
agreement with experimentally observed rotational bands.

\section{Rotational invariance of the su$_q$(2) Hamiltonian} 

In this section we are going to use both the usual quantum mechanical 
operators of angular momentum, denoted by $\hat l_+$, $\hat l_-$, $\hat l_0$, 
and the $q$-deformed ones, which are related to su$_q$(2) and denoted by 
$\hat L_+$, $\hat L_-$, $\hat L_0$, as in Sec. 2. In this section we are 
going to use hats (\ $\hat {}$\ ) for the operators, in order to give emphasis 
to the distinction between the operators and their eigenvalues. 
For brevity we are going 
to call the operators $\hat l_+$, $\hat l_-$, $\hat l_0$ {\sl ``classical''}, 
while the operators $\hat L_+$, $\hat L_-$, $\hat L_0$ will be called 
{\sl ``quantum''}. For the ``classical''
basis the symbol $|l,{\rm m}\rangle _c$ will be used, while the 
``quantum'' basis will be denoted by $|\ell, m\rangle _q$, as in Sec. 2. 
Therefore $l$ and ${\rm m}$
are the quantum numbers related to the usual quantum mechanical angular 
momentum, which is characterized by the su(2) symmetry, 
while $\ell$ and $ m$ are the quantum numbers related 
to the deformed angular momentum, which is characterized by the su$_q$(2) 
symmetry. 

The ``classical'' operators satisfy the usual su(2) commutation relations
\begin{equation} \label{eq:2.1}
[\hat l_0, \hat l_{\pm}]= \pm \hat l_{\pm}, \qquad [\hat l_+, \hat l_-]=2
\hat l_0,
\end{equation}
while the finite-dimensional irreducible representation $D^l$ of su(2) 
is determined by the highest weight vector $| l, l\rangle _c $ with
\begin{equation} \label{eq:2.2}
\hat l_+|l,l\rangle _c = 0, 
\end{equation} 
and the basis states $|l,{\rm m}\rangle _c$ are expressed as 
\begin{equation} \label{eq:2.3}
|l,{\rm m}\rangle _c = \sqrt{ (l+{\rm m})! \over (2l)! (l-{\rm m})!} 
(\hat l_-)^{l-{\rm m}} |l,l\rangle _c. 
\end{equation}
The action of the generators of su(2) on the vectors of the ``classical'' 
basis is described by
\begin{equation} \label{eq:2.4}
\hat l_{\pm} | l, {\rm m}\rangle _c = \sqrt{(l\mp {\rm m})(l\pm {\rm m}+1)} 
|l, {\rm m}\pm 1\rangle _c, 
\end{equation} 
\begin{equation} \label{eq:2.5}
\hat l_0 |l,{\rm m}\rangle _c = {\rm m} |l,{\rm m}\rangle _c,
\end{equation}
the dimension of the corresponding representation being dim$D^l=2l+1$ 
for $l=0$, ${1\over 2}$, 1, ${3\over 2}$, 2, \dots

The second order Casimir operator of su(2) is 
\begin{equation} \label{eq:2.6} 
\hat C_2 = {1\over 2} (\hat l_+ \hat l_-+\hat l_-\hat l_+) + \hat l_0^2 = 
\hat l_- \hat l_+ +\hat l_0(\hat l_0+1) 
= \hat l_+ \hat l_- +\hat l_0(\hat l_0-1),  
\end{equation}
where the symbol $1$ is used for the unit operator, 
while its eigenvalues in the space of the irreducible representation $D^l$ 
are $l(l+1)$
\begin{equation} \label{eq:2.7} 
\hat C_2|l, {\rm m}\rangle _c = l(l+1) |l,{\rm m}\rangle _c.
\end{equation}
It is useful to introduce the operator $\hat l$ through the definition 
\begin{equation} \label{eq:2.8} 
\hat C_2 \equiv \hat l(\hat l+1).
\end{equation}
Insisting that $\hat l$ should be a positive operator one then has by solving 
the relevant quadratic equation and keeping only the positive sign in front 
of the square root \cite{CGZ}
\begin{equation} \label{eq:2.9} 
\hat l= {1\over 2} (-1+\sqrt{1+4 \hat C_2})
\end{equation}
The action of the operator $\hat l$ on the vectors of the ``classical'' basis 
is then given by 
$$\hat l|l, {\rm m}\rangle_c ={1\over 2} (-1+\sqrt{1+4 \hat C_2}) 
|l, {\rm m}\rangle _c 
= {1\over 2} (-1+\sqrt{1+4 l(l+1)}) |l, {\rm m}\rangle _c $$
\begin{equation} \label{eq:2.10} 
={1\over 2} (-1+\sqrt{(2l+1)^2}) |l, {\rm m}\rangle _c 
= {1\over 2} (-1+2l+1) |l, {\rm m}\rangle_c = l | l, {\rm m}\rangle_c,  
\end{equation}
where again only the positive value of the square root has been taken 
into account. 

In this ``classical'' environment one can introduce the operators 
\cite{CGZ,CZPLB} 
\begin{equation} \label{eq:2.11}
\hat {\cal L}_+= \sqrt{ [\hat l+\hat l_0] [\hat l-\hat l_0+1] 
\over (\hat l+\hat l_0) (\hat l-\hat l_0+1)}  \hat l_+, 
\end{equation}
\begin{equation} \label{eq:2.12}
\hat {\cal L}_-= \hat l_-\sqrt{ [\hat l+\hat l_0] [ \hat l-\hat l_0+1] 
\over (\hat l+\hat l_0)( \hat l-\hat l_0+1)}, 
\end{equation}
\begin{equation} \label{eq:2.13}
\hat {\cal L}_0=\hat l_0,
\end{equation}
where square brackets denote $q$-operators, as defined in Eq. (\ref{eq:1.3}).  

The action of these operators on the vectors of the ``classical'' basis 
is given by 
$$\hat {\cal L}_+ |l, {\rm m}\rangle_c = \sqrt{ [\hat l+\hat l_0]
[\hat l-\hat l_0+1] \over (\hat l+\hat l_0)(\hat l-\hat l_0+1)} 
\hat l_+|l, {\rm m}\rangle _c $$
$$= \sqrt{[\hat l+\hat l_0][\hat l-\hat l_0+1]\over(\hat l+\hat l_0)
(\hat l-\hat l_0+1)}
\sqrt{(l-{\rm m})(l+{\rm m}+1)} |l, {\rm m}+1\rangle _c $$
$$= \sqrt{[l+{\rm m}+1][l-{\rm m}]\over (l+{\rm m}+1)(l-{\rm m})}
\sqrt{(l-{\rm m})(l+{\rm m}+1)} |l, {\rm m}+1\rangle _c $$
\begin{equation} \label{eq:2.14}
= \sqrt{[l+{\rm m}+1][l-{\rm m}]} |l, {\rm m}+1\rangle_c  ,
\end{equation} 
$$ \hat {\cal L}_-|l,{\rm m}\rangle_c = \hat l_- \sqrt{ [\hat l+\hat l_0]
[\hat l-\hat l_0+1]\over (\hat l+\hat l_0)(\hat l-\hat l_0+1)} 
|l, {\rm m}\rangle _c 
=\hat l_- \sqrt{[l+{\rm m}][l-{\rm m}+1]\over (l+{\rm m})
(l-{\rm m}+1)} |l, {\rm m}\rangle _c $$
\begin{equation} \label{eq:2.15} 
=  \sqrt{ (l+{\rm m})(l-{\rm m}+1)} \sqrt{[l+{\rm m}][l-{\rm m}+1]\over 
(l+{\rm m})(l-{\rm m}+1)} |l, {\rm m}-1\rangle _c 
= \sqrt{[l+{\rm m}][l-{\rm m}+1]} |l, {\rm m}-1\rangle _c, 
\end{equation}
\begin{equation} \label{eq:2.16} 
\hat {\cal L}_0 |l, {\rm m}\rangle_c = \hat l_0 |l, {\rm m}\rangle_c 
= m |l, {\rm m}\rangle _c, 
\end{equation}
or, in compact form, 
\begin{equation} \label{eq:2.17} 
\hat {\cal L}_{\pm} |l, {\rm m}\rangle _c = \sqrt{[l\mp {\rm m}][l\pm 
{\rm m}+1]} |l, {\rm m} \pm 1\rangle_c , \qquad \hat {\cal L}_0 |l, 
{\rm m}\rangle_c = {\rm m} |l, {\rm m}\rangle _c . 
\end{equation}

It is clear that the operators $\hat {\cal L}_+$ and $\hat l_+$ do not commute 
$$[\hat {\cal L}_+, \hat l_+]|l, {\rm m}\rangle_c = \hat {\cal L}_+ \hat l_+ 
|l, {\rm m}\rangle_c - \hat l_+ \hat {\cal L}_+ | l, {\rm m}\rangle _c$$
$$= \hat {\cal L}_+ \sqrt{ (l-{\rm m})(l+{\rm m}+1)} |l, {\rm m}+1\rangle_c 
-\hat l_+ \sqrt{[l-{\rm m}][l+{\rm m}+1]} |l, {\rm m}+1\rangle_c $$
$$= (\sqrt{[l-{\rm m}-1][l+{\rm m}+2]} \sqrt{(l-{\rm m})(l+{\rm m}+1)} $$
\begin{equation} \label{eq:2.31} 
-\sqrt{(l-{\rm m}-1)(l+{\rm m}+2)} \sqrt{[l-{\rm m}][l+{\rm m}+1]})
|l, {\rm m}+2\rangle _c \neq 0.
\end{equation}
This result is expected if one considers Eq. (\ref{eq:2.11}): The operator
$\hat l_+$  does commute with itself and with the operator $\hat l$, 
which is a function 
of the relevant Casimir operator, as Eq. (\ref{eq:2.9}) indicates, but it 
does not commute with the operator $\hat l_0$, as Eq. (\ref{eq:2.1}) shows. 
In the same way one can see that 
\begin{equation} \label{eq:2.32}
[\hat {\cal L}_-, \hat l_-]|l, {\rm m}\rangle _c \neq 0. 
\end{equation}

One can now prove that the ``new'' operators satisfy the commutation 
relations of Eqs. (\ref{eq:1.1}), (\ref{eq:1.2}). Indeed one has 
$$ [\hat {\cal L}_0,\hat {\cal L}_+]| l, {\rm m}\rangle_c = \hat {\cal L}_0 
\hat {\cal L}_+ 
|l,{\rm m}\rangle _c - \hat {\cal L}_+ \hat {\cal L}_0 |l, {\rm m}\rangle _c$$
$$= \hat {\cal L}_0 \sqrt{[l-{\rm m}] [l+{\rm m}+1]} |l, {\rm m}+1\rangle_c- 
\hat {\cal L}_+ m |l, {\rm m}\rangle _c $$
$$ = ({\rm m}+1) \sqrt{[l-{\rm m}][l+{\rm m}+1]} |l, {\rm m}+1\rangle _c 
-\sqrt{[l-{\rm m}][l+{\rm m}+1]} {\rm m} |l, {\rm m}+1\rangle _c $$
\begin{equation} \label{eq:2.18} 
= ({\rm m}+1-{\rm m}) \sqrt{[l-{\rm m}][l+{\rm m}+1]} |l, {\rm m}+1\rangle_c 
= \hat {\cal L}_+|l, {\rm m}\rangle _c. 
\end{equation}
and in exactly the same way 
\begin{equation} \label{eq:2.19} 
[\hat {\cal L}_0, \hat {\cal L}_-] |l, {\rm m}\rangle _c = -\hat 
{\cal L}_-|l, {\rm m}\rangle_c, 
\end{equation}
while for the commutator of Eq. (\ref{eq:1.2}) one has 
$$[\hat {\cal L}_+, \hat {\cal L}_-] |l, {\rm m}\rangle _c = \hat {\cal L}_+ 
\hat {\cal L}_- 
|l, {\rm m}\rangle _c -\hat {\cal L}_-\hat {\cal L}_+|l, {\rm m}\rangle _c $$
$$= \hat {\cal L}_+ \sqrt{[l+{\rm m}][l-{\rm m}+1]} |l, {\rm m}-1\rangle _c 
-\hat {\cal L}_- \sqrt{[l-{\rm m}][l+{\rm m}+1]} |l, {\rm m}+1\rangle _c $$
\begin{equation} \label{eq:2.20} 
= ([l+{\rm m}][l-{\rm m}+1]-[l-{\rm m}][l+{\rm m}+1]) |l, {\rm m}\rangle _c 
= [2{\rm m}] |l, {\rm m}\rangle _c = [2\hat {\cal L}_0] |l, {\rm m}\rangle _c ,
\end{equation}
where use of the identity 
\begin{equation} \label{eq:2.21} 
[l+{\rm m}][l-{\rm m}+1] -[l-{\rm m}] [l+{\rm m}+1] = [2{\rm m}], 
\end{equation}
which can be easily proved by using Eq. (\ref{eq:1.3}), has been made. 

We have therefore demonstrated that the operators $\hat {\cal L}_+$,
$\hat {\cal L}_-$, $\hat {\cal L}_0$ satisfy the commutation relations 
of the su$_q$(2) algebra. As a consequence, the quantities appearing 
in the rhs of Eqs. (\ref{eq:2.11})-(\ref{eq:2.13}) are just the realizations
of the generators of su$_q$(2) in the ``classical'' basis. Therefore 
from now on we can use the symbols $\hat L_+$, $\hat L_-$, $\hat L_0$ 
in the place of $\hat {\cal L}_+$, $\hat {\cal L}_-$, $\hat {\cal L}_0$.  

One can also see that the operator 
\begin{equation} \label{eq:2.22}
\hat C= \hat L_-\hat L_++[\hat L_0][\hat L_0+1]
\end{equation}
acts on the vectors of the ``classical'' basis  as 
$$\hat C|l,{\rm m}\rangle_c = \hat L_-\hat L_+|l, {\rm m}\rangle _c + 
[\hat L_0][\hat L_0+1] |l, {\rm m}\rangle _c $$
$$= \hat L_- \sqrt{[l+{\rm m}][l-{\rm m}+1]} |l, {\rm m}+1\rangle _c 
+ [{\rm m}] [{\rm m}+1] |l, {\rm m}\rangle _c $$
\begin{equation} \label{eq:2.23} 
= [l+{\rm m}] [l-{\rm m}+1] |l, {\rm m}\rangle _c + [{\rm m}] [{\rm m}+1] 
|l, {\rm m}\rangle _c = [l] [l+1] |l, {\rm m}\rangle_c,
\end{equation}
where in the last step the identity 
\begin{equation} \label{eq:2.24} 
[l+{\rm m}] [l-{\rm m}+1] + [m] [m+1] = [l] [l+1], 
\end{equation}
which can easily be verified using Eq. (\ref{eq:1.3}), has been used. 

Using Eqs. (\ref{eq:2.17}), (\ref{eq:2.23}) 
one can now prove that the operator $\hat C$ commutes with 
the generators $\hat L_+$, $\hat L_-$, $\hat L_0$ of su$_q$(2), 
i.e. that $\hat C$ is the second 
order Casimir operator of su$_q$(2). Indeed one has 
$$[\hat C, \hat L_+]|l, {\rm m}\rangle_c = \hat C \hat L_+|l, {\rm m}\rangle_c
- \hat L_+ \hat C |l, {\rm m}\rangle_c $$
$$= \hat C \sqrt{[l-{\rm m}][l+{\rm m}+1]} |l, {\rm m}+1\rangle _c -
\hat L_+[l][l+1] |l, {\rm m}\rangle _c $$
\begin{equation} \label{eq:2.25}
=[l][l+1] \sqrt{[l-{\rm m}][l+{\rm m}+1]} |l, {\rm m}+1\rangle _c 
-\sqrt{[l-{\rm m}][l+{\rm m}+1]} [l][l+1] |l, {\rm m}\rangle _c =0.
\end{equation}
In exactly the same way one can prove that 
\begin{equation} \label{eq:2.26}
[\hat C, \hat L_-] |l, {\rm m}\rangle_c =0, 
\end{equation}
while in addition one has 
$$[\hat C, \hat L_0] |l, {\rm m}\rangle_c = \hat C \hat L_0 
|l, {\rm m}\rangle_c -\hat L_0 \hat C |l, {\rm m}\rangle _c $$
\begin{equation} \label{eq:2.27}
= [l][l+1] {\rm m} |l, {\rm m}\rangle_c - m [l][l+1] |l, {\rm m}\rangle_c =0. 
\end{equation}

Thus we have proved that the operator $\hat C$ is the second order Casimir 
operator of su$_q$(2). We are now going to prove that the operator $\hat C$ 
commutes also with the generators $\hat l_+$, $\hat l_-$, $\hat l_0$ 
of the usual su(2) algebra. Indeed one has 
$$[\hat C, \hat l_+] |l, {\rm m}\rangle_c = \hat C \hat l_+ 
|l, {\rm m}\rangle_c -\hat l_+ \hat C |l, {\rm m}\rangle _c $$
$$= \hat C \sqrt{(l-{\rm m})(l+{\rm m}+1)} |l, {\rm m}+1\rangle_c 
-\hat l_+[l][l+1] |l, {\rm m}\rangle _c $$
\begin{equation} \label{eq:2.28}
= [l][l+1] \sqrt{(l-{\rm m})(l+{\rm m}+1)} |l, {\rm m}+1\rangle _c 
-\sqrt{(l-{\rm m})(l+{\rm m}+1)} [l][l+1] |l, {\rm m}+1\rangle _c =0.
\end{equation}
In exactly the same way one can prove that 
\begin{equation} \label{eq:2.29} 
[\hat C, \hat l_-] |l, {\rm m}\rangle _c =0, 
\end{equation}
while the relation
\begin{equation} \label{eq:2.30} 
[\hat C, \hat l_0] |l, {\rm m}\rangle _c =0 
\end{equation}
occurs from Eq. (\ref{eq:2.27}), since $\hat L_0=\hat l_0$ by definition 
(see Eq. (\ref{eq:2.13})~).

The following comments are now in place: 
 
a) The fact that the operator $\hat C$, which will be from now on denoted by 
$\hat C_2^{(q)}$, commutes with the generators of su(2) implies that 
this operator is a function of the second order Casimir operator of 
su(2), given in Eq. (\ref{eq:2.6}). 
As a consequence, it should be possible to express
the eigenvalues of $\hat C_2^{(q)}$, which are $[l][l+1]$  (as we have seen in 
Eq. (\ref{eq:2.23})~), in terms of the eigenvalues of $\hat C_2$, which 
are $l(l+1)$ (as we have seen in Eq. (\ref{eq:2.7})~). 
This task will be undertaken in the next section. 

b) Eqs. (\ref{eq:2.28})-(\ref{eq:2.30}) 
also tell us that the Hamiltonian of Eq. (\ref{eq:1.13}) 
commutes with the generators of the usual su(2) algebra, i.e. it is 
rotationally invariant. The Hamiltonian of Eq. (\ref{eq:1.13}) does not 
break rotational symmetry. It corresponds to a function of the second order 
Casimir operator of the usual su(2) algebra. This function, however, 
has been chosen in an appropriate way, in order to guarantee that 
the Hamiltonian of Eq. (\ref{eq:1.13}) is also invariant under a more
complicated symmetry, namely the symmetry su$_q$(2). 

c) From the contents of the present section it is also clear that the irrep
$D^{\ell}_{(q)}$ of su$_q$(2) and the irrep $D^l$ of su(2) have the same 
structure, the relevant states being in an one to one correspondence to each 
other. The similarity between Eqs. (\ref{eq:2.17}) and 
(\ref{eq:2.4})-(\ref{eq:2.5}) implies that the distinction 
between the ``classical'' basis of the present section 
and the ``quantum'' basis of Section 2 turns out to 
be unnecessary, as well as that the quantum numbers $\ell$ and $m$ 
can be identified with the usual angular momentum quantum numbers 
$l$ and ${\rm m}$.  

d) These conclusions are valid in the case of $q$ being not a root of unity,
as already mentioned in Sec. 2. 

\section{Exact expansion of the su$_q$(2) spectrum}

Let us consider the spectrum of Eq. (\ref{eq:1.17}), which has been found 
relevant 
to rotational nuclear and molecular spectra, assuming for simplicity $E_0=0$
and $\tau > 0$. Since the Hamiltonian of Eq. (\ref{eq:1.13}) is invariant 
under su(2), as we have seen in the previous section, it should be possible 
in principle to express it as a function of the Casimir operator $C_2$ 
of the usual su(2) algebra. As a consequence, it should also be possible 
to express the eigenvalues of this Hamiltonian, given in Eq. (\ref{eq:1.17}),
as a function of the eigenvalues of the Casimir operator of the usual 
su(2), i.e. as a function of $\ell(\ell+1)$.    
This is a nontrivial task, since in Eq. (\ref{eq:1.17}) two different 
functions of the variable $\ell$ appear, while we are in need of a single 
function of the variable $\ell(\ell+1)$, which is related to the length 
of the angular momentum vector.  
In order to represent the expression of Eq. (\ref{eq:1.17}) 
as a power series of the variable $\ell(\ell+1)$, one can use the identity
\begin{equation} \label{eq:3.1}
\sin(\ell\tau)\sin((\ell+1)\tau) 
=\frac{1}{2}\left\{\cos(\tau)-\cos((2\ell+1)\tau)\right\}.
\end{equation} %
It turns out that the coefficients of the relevant expansion can be expressed 
in terms of the spherical Bessel functions of  the first kind $j_n(x)$ 
\cite{AbrSte}, which are determined through the generating function 
\begin{equation} \label{eq:3.2}
{1\over x} \cos\sqrt{x^2-2xt} = \sum_{n=0}^\infty j_{n-1}(x) {t^n \over n!},
\end{equation} %
and are characterized by the asymptotic behavior
\begin{equation} \label{eq:3.3} 
j_n(x)\approx\frac{x^n}{(2n+1)!!}, \qquad x\ll 1 .
\end{equation}
Performing the substitutions
\begin{equation} \label{eq:3.4} 
x=\tau, \qquad t=-2\tau \ell (\ell+1) ,
\end{equation}  
which imply 
\begin{equation} \label{eq:3.5}  
x^2-2xt=\tau^2(2\ell+1)^2,  
\end{equation}
one gets the expression
\begin{equation} \label{eq:3.6} 
\frac{1}{\tau}\cos((2\ell+1)\tau)=
\sum_{n=0}^{\infty}\frac{(-2\tau)^n}{n!}\,j_{n-1}(\tau)\,\{\ell(\ell+1)\}^n, 
\end{equation} %
which in the special case of $\ell=0$ reads
\begin{equation} \label{eq:3.7} 
{1\over \tau} \cos\tau = j_{-1}(\tau),
\end{equation}
in agreement with the definition  \cite{AbrSte}
\begin{equation} \label{eq:3.8} 
j_{-1}(x) = {\cos x \over x}.
\end{equation}
Substituting Eqs. (\ref{eq:3.6}) and (\ref{eq:3.7}) in Eq. (\ref{eq:3.1}),
and taking into account that \cite{AbrSte}
\begin{equation} \label{eq:3.9}
j_0(x) = {\sin x \over x},
\end{equation} %
Eq. (\ref{eq:1.17}) takes the form 
\begin{equation} \label{eq:3.10} 
E_{\ell}^{(\tau)}=\frac{A}{j_0^2(\tau)}\sum_{n=0}^{\infty}
\frac{(-1)^n(2\tau)^n}{(n+1)!}\,j_n(\tau)\,\{\ell(\ell+1)\}^{n+1}, 
\end{equation} %
which is indeed an expansion in terms of $\ell(\ell+1)$. 

\section{Approximate expansion of the su$_q$(2) spectrum} 

We are now going to consider an approximate form of this expansion, which 
will allow us to connect the present approach to the description of nuclear
spectra proposed by Amal'sky \cite{Amal}. 

For ``small deformation'', i.e. for $\tau\ll1$, one can use the asymptotic 
expression of Eq. (\ref{eq:3.3}). 
Keeping only the terms of the lowest order one then 
obtains the following approximate series
\begin{equation} \label{eq:4.1}
E_{\ell}^{(\tau)}\approx A\,\sum_{n=0}^{\infty}
\frac{(-1)^n(2\tau)^{2n}}{(n+1)(2n+1)!}\,\{\ell(\ell+1)\}^{n+1}, 
\end{equation} %
where use of the identity 
\begin{equation} \label{eq:4.2}
2^n (n+1)! (2n+1)!! = (n+1) (2n+1)!
\end{equation} %
has been made. 
The first few terms of this expansion are
\begin{equation} \label{eq:4.3} 
E_{\ell}^{(\tau)}\approx A\Bigl(\,\ell(\ell+1)
-\frac{\tau^2}{3}\{\ell(\ell+1)\}^2 +\frac{2\tau^4}{45}\{\ell(\ell+1)\}^3
-\frac{\tau^6}{315}\{\ell(\ell+1)\}^4+\ldots\Bigr),
\end{equation} %
in agreement with the findings of Ref. \cite{PLB251}. 

One can now observe that the expansion appearing in Eq. (\ref{eq:4.1})   
is similar to the power series of the function
\begin{equation} \label{eq:4.4} 
\sin^2 x=\frac{1}{2}(1-\cos 2x)=
\sum_{k=1}^{\infty}(-1)^{k+1}2^{2k-1}\frac{x^{2k}}{(2k)!}.
\end{equation} %
Then, performing the auxiliary substitution
\begin{equation} \label{eq:4.5} 
\xi=\sqrt{\ell(\ell+1)}, \qquad\qquad \eta=\ell(\ell+1)=\xi^2, 
\end{equation} %
one can put the expansion of Eq. (\ref{eq:4.1}) in the form 
\begin{equation} \label{eq:4.6} 
E_\ell^{(\tau)} \approx A\,\frac{\sin^2(\tau\xi)}{\tau^2}=
\frac{\hbar^2}{2 {\cal J}_0}\frac{\sin^2(\tau\sqrt{\ell(\ell+1)})}{\tau^2},
\qquad q=e^{i\tau}.  
\end{equation}  %
This result is similar to the expression proposed for the unified 
description of nuclear rotational spectra by G. Amal'sky \cite{Amal}
\begin{equation} \label{eq:4.7}
E_\ell=\varepsilon_0\,\sin^2\left(\frac{\pi}{N}\sqrt{\ell(\ell+1)}\right), 
\end{equation}
where $\varepsilon_0$ is a phenomenological constant 
($\varepsilon_0\approx 6.664$ MeV) which remains the same for all nuclei, 
while $N$ is a free parameter varying from one nucleus to the other. 

\section{Analytic expressions based on the approximate expansion
of the su$_q$(2) spectrum}

In this  section we will consider some analytic expressions, which are based 
on the approximate result of Eq. (\ref{eq:4.6}), with the purpose 
of connecting the present approach to the Harris formalism \cite{Harris}. 
In the study of high spin phenomena the rotational frequency $\omega$ 
and the kinematic moment of inertia ${\cal J}$ are defined by 
\begin{equation} \label{eq:5.1}
\hbar\omega=\frac{\partial E}{\partial \xi},
\end{equation}
\begin{equation} \label{eq:5.2} 
\frac{\hbar^2}{2{\cal J}}=\frac{\partial E}{\partial \eta}=
\frac{1}{2\xi}\frac{\partial E}{\partial \xi}, 
\end{equation}
where $\xi$ has been defined in Eq. (\ref{eq:4.5}) and 
\begin{equation} \label{eq:5.3} 
\eta = \ell(\ell+1) = \xi^2. 
\end{equation}
From Eqs. (\ref{eq:5.1}) and (\ref{eq:5.2}) it is clear that the two quantities
are connected by the relation
\begin{equation} \label{eq:5.4} 
{\cal J} \omega = \hbar \xi.
\end{equation}
Applying these definitions to the analytical
expression of Eq. (\ref{eq:4.6}) one obtains
\begin{equation} \label{eq:5.5}
\hbar \omega= A {\sin(2\tau\xi)\over \tau} = 
\frac{\hbar^2}{2{\cal J}_0}\frac{\sin(2\tau\xi)}{\tau},
\end{equation}
\begin{equation} \label{eq:5.6} 
{\cal J}={\cal J}_0\,\frac{2\tau\xi}{\sin(2\tau\xi)},
\end{equation}
where the identity 
\begin{equation} \label{eq:5.7}
\sin 2x= 2\sin x \cos x
\end{equation}
has been used. 
Using the expressions for $E$ (for which we drop the superscript and subscript)
and $\omega$ given in Eqs. (\ref{eq:4.6}) and (\ref{eq:5.5}) 
one can easily verify that 
\begin{equation} \label{eq:5.8}
{ {\cal J}_0 \omega^2 \over 2} = E \left( 1-{\tau^2 \over A} E\right),
\end{equation}
where use of the identity of Eq. (\ref{eq:5.7}) has been made. 
Defining 
\begin{equation} \label{eq:5.9} 
\varepsilon= {\tau^2 \over A} E = \sin^2(\tau\xi),
\end{equation}
\begin{equation} \label{eq:5.10} 
t={\hbar \tau \over A} = {2 {\cal J}_0 \over \hbar} \tau,
\end{equation}
where $t$ is a constant possessing dimensions of time, 
Eq. (\ref{eq:5.8}) takes the form 
\begin{equation} \label{eq:5.11}
(\omega t)^2 = 4\varepsilon(1-\varepsilon) = 4\varepsilon -4\varepsilon^2. 
\end{equation}
This expression can be considered as a quadratic equation for $\varepsilon$, 
allowing us to express $\varepsilon$ as a function of $\omega t$. Indeed 
one finds 
\begin{equation}\label{eq:5.12} 
\varepsilon= {1\over 2} ( 1\pm \sqrt{1-(\omega t)^2}).
\end{equation}
Using the Taylor expansion \cite{AbrSte}
\begin{equation} \label{eq:5.13}
(1+x)^{-1/2} = 1 -{1\over 2} x + {1 \cdot 3 \over 2\cdot 4} x^2 -
{1\cdot 3 \cdot 5 \over 2 \cdot 4 \cdot 6} x^3 +\ldots, \qquad 
-1<x\leq 1
\end{equation}
one obtains 
\begin{equation} \label{eq:5.14} 
\varepsilon = {1\over 2} \left(1\pm\left(1-{1\over 2} (\omega t)^2 
- {1\over 8} (\omega t)^4 -{5\over 16} (\omega t)^6 -\ldots    
\right)\right)
\end{equation}
The choice of the negative sign then leads to
\begin{equation} \label{eq:5.15} 
\varepsilon = {1\over 4} (\omega t)^2 + {1\over 16} (\omega t)^4 
+{5\over 32} (\omega t)^6+\ldots,
\end{equation} 
which through Eq. (\ref{eq:5.9}) gives
\begin{equation} \label{eq:5.16}
E = {A\over (2 \tau)^2} \left(  (\omega t)^2 + {1\over 4} (\omega t)^4 
+{5\over 8} (\omega t)^6 +\ldots \right). 
\end{equation}
The choice of the positive sign gives correspondingly  
\begin{equation} \label{eq:5.17} 
\varepsilon= 1 - {1\over 4} (\omega t)^2 - {1\over 16} (\omega t)^4 
-{5\over 32} (\omega t)^6 -\ldots 
\end{equation}
and 
\begin{equation} \label{eq:5.18}
E = {A\over \tau^2} \left(  1 - {1\over 4} (\omega t)^2 - {1\over 16}
(\omega t)^4 - {5\over 32} (\omega t)^6 -\ldots \right)
\end{equation}

It is clear that Eq. (\ref{eq:5.16}) corresponds to $E$ increasing as a 
function of $\omega$, while Eq. (\ref{eq:5.18}) corresponds to $E$
decreasing as a function of $\omega$. Therefore only the first solution 
can be relevant to the description of nuclear rotational spectra. 

We are now trying to find a similar expansion for the kinematic moment 
of inertia ${\cal J}$. Using Eq. (\ref{eq:5.10}) 
one can rewrite Eq. (\ref{eq:5.5}) in the form
\begin{equation} \label{eq:5.19}
\omega t = \sin(2\tau\xi). 
\end{equation} 
Then Eq. (\ref{eq:5.6}) gives 
\begin{equation} \label{eq:5.20}
{ {\cal J} \over {\cal J}_0} = {2\tau\xi \over \sin(2\tau\xi)}
={ {\rm arcsin}(\omega t) \over \omega t}.
\end{equation}
Using then the Taylor expansion \cite{AbrSte}
\begin{equation} \label{eq:5.21} 
{\rm arcsin}x = x + {1\over 2} {x^3 \over 3} + {1\cdot 3 \over 2\cdot 4} 
{x^5 \over 5} + {1\cdot 3 \cdot 5\over 2 \cdot 4 \cdot 6} {x^7 \over 7} 
+\ldots 
\end{equation}
one obtains 
\begin{equation} \label{eq:5.22}
{ {\cal J} \over {\cal J}_0} = 1 + {1\over 6} (\omega t)^2 + {3\over 40} 
(\omega t)^4 + {5\over 112} (\omega t)^6 +\ldots 
\end{equation}
Using Eqs. (\ref{eq:5.4}) and (\ref{eq:5.10}) one finds from this result that 
\begin{equation} \label{eq:5.23} 
\xi =\sqrt{\ell(\ell+1)} = {\omega {\cal J} \over \hbar} =
{1\over 2\tau} \left( \omega t + {1\over 6} (\omega t)^3 + {3\over 40}
(\omega t)^5 + {5\over 112} (\omega t)^7 +\ldots \right) 
\end{equation}

The expansions appearing in Eqs. (\ref{eq:5.16}) and (\ref{eq:5.23}) 
are of the form occuring in the Harris formalism \cite{Harris} 
\begin{equation} \label{eq:5.24}
E= E_0+ {1\over 2} ({\cal J}_0 \omega^2 + 3 C \omega^4 + 5 D \omega^6 
+ 7 F \omega^8 +\ldots), 
\end{equation}
\begin{equation} \label{eq:5.25} 
\sqrt{\ell(\ell+1)} = {\cal J}_0 \omega + 2 C \omega^3 + 3 D \omega^5 
+ 4 F \omega^7 +\ldots, 
\end{equation}
the main difference between the two formalisms being the fact that in the case 
of Harris the coefficients of the various terms in the series are independent 
from each other, while in the present case the coefficients in the series are
interdependent, since they all contain the constant $t$. 
It should be noticed at this point that the Harris formalism is known
\cite{KDD333} 
to be equivalent to the Variable Moment of Inertia (VMI) model \cite{VMI}. 
The similarities between the su$_q$(2) approach and the VMI model have been 
directly considered in Ref. \cite{PLB251}. 

\section{Irreducible tensor operators under su$_q$(2)}

A different path towards the construction of a Hamiltonian appropriate 
for the description of rotational spectra can be taken through the 
construction of irreducible tensor operators under su$_q$(2)
\cite{STK593,STK690}. In this discussion we limit ourselves to real values 
of $q$, i.e. to $q=e^\tau$ with $\tau$ being real, as in Refs. 
\cite{STK593,STK690}.  

An irreducible tensor operator of rank $k$ is the 
set of $2k+1$ operators $T^{(q)}_{k,\kappa}$ ($\kappa=k$, $k-1$, $k-2$, 
$\ldots$, $-k$), which satisfy with the generators of the su$_q$(2) 
algebra the commutation relations \cite{STK593,STK690}  
\begin{equation} \label{eq:6.1}
[L_0, T^{(q)}_{k,\kappa}]= \kappa T^{(q)}_{k,\kappa},
\end{equation}
\begin{equation} \label{eq:6.2}
[L_{\pm}, T^{(q)}_{k,\kappa}]_{q^\kappa} = \sqrt{[k\mp \kappa] [k\pm \kappa+1]}
T^{(q)}_{k,\kappa\pm 1} q^{-L_0},
\end{equation}
where $q$-commutators are defined by  
\begin{equation} \label{eq:6.3} 
[A,B]_{q^\alpha} = A B-q^\alpha B A.
\end{equation}
It is clear that in the limit $q\rightarrow 1$ these commutation relations 
reduce to the usual ones, which occur in the definition of irreducible 
tensor operators under su(2). 
It should also be noticed that the operators
\begin{equation} \label{eq:6.4} 
R^{(q)}_{k,\kappa} = (-1)^\kappa q^{-\kappa} (T^{(q)}_{k,-\kappa})^\dagger,
\end{equation}
where $\dagger$ denotes Hermitian conjugation, satisfy the same commutation
relations (\ref{eq:6.1}), (\ref{eq:6.2}) as the operators $T^{(q)}_{k,\kappa}$,
i.e. the operators $R^{(q)}_{k,\kappa}$ also form an irreducible tensor 
operator of rank $k$ under su$_q$(2). 

We can construct an irreducible tensor operator of rank 1 using as building 
blocks the generators of su$_q$(2). This irreducible tensor operator will
consist of the operators $J_{+1}$, $J_{-1}$, $J_0$, which should satisfy 
the commutation relations 
\begin{equation} \label{eq:6.5} 
[L_0, J_m]= m J_m,
\end{equation}
\begin{equation} \label{eq:6.6} 
[L_{\pm}, J_m ]_{q^m} = \sqrt{[1\mp m] [2\pm m]} J_{m\pm 1} q^{-L_0},
\end{equation}
which are a special case of Eqs. (\ref{eq:6.1}), (\ref{eq:6.2}), while 
the relevant Hermitian conjugate operators will be 
\begin{equation} \label{eq:6.7} 
(J_m)^\dagger = (-1)^m q^{-m} J_{-m},
\end{equation} 
which is a consequence of Eq. (\ref{eq:6.4}). 
It turns out \cite{STK593,STK690,JPA6939}
that the explicit form of the relevant operators is
\begin{equation} \label{eq:6.8}
J_{+1}= -{1\over \sqrt{[2]}} q^{-L_0} L_+, 
\end{equation}
\begin{equation} \label{eq:6.9}
J_{-1}={1\over \sqrt{[2]}} q^{-L_0} L_-,
\end{equation} 
$$ J_0={1\over [2]} (qL_+ L_- -q^{-1}L_- L_+) = {1\over [2]} (q L_- L_+ 
- q^{-1} L_+ L_-) +[2L_0] $$
$$= {1\over 2} \left( [2L_0]+{(q-q^{-1})\over [2]} (L_- L_+ + L_+ L_-) \right) 
=  {1\over [2]} (q [2L_0] + (q-q^{-1}) L_- L_+) $$
\begin{equation} \label{eq:6.10}
={1\over [2]} \left( q [2L_0]+(q-q^{-1}) ( C_2^{(q)}-[L_0][L_0+1])\right),
\end{equation}
while the Hermitian conjugate operators are 
\begin{equation} \label{eq:6.11}
(J_{+1})^\dagger =-q^{-1} J_{-1}, \quad (J_{-1})^\dagger = -q J_{+1}, \quad
(J_0)^\dagger = J_0. 
\end{equation}
It is clear that in the limit $q\rightarrow 1$ these results reduce to the 
usual expressions for spherical tensors of rank 1 under su(2), formed out 
of the usual angular momentum operators 
\begin{equation} \label{eq:6.12}
J_+=-{L_+\over \sqrt{2}}=-{L_x+iL_y \over \sqrt{2}}, \qquad 
J_-={L_-\over \sqrt{2}}={L_x-iL_y\over \sqrt{2}}, \qquad J_0 = L_0,
\end{equation} 
\begin{equation} \label{eq:6.13}
(J_+)^\dagger = -J_-, \quad (J_-)^\dagger = -J_+, \quad (J_0)^\dagger =J_0. 
\end{equation}

The commutation relations among the operators $J_{+1}$, $J_{-1}$, $J_0$ can be 
obtained using Eqs. (\ref{eq:6.8})-(\ref{eq:6.10}) and (\ref{eq:6.5}), 
(\ref{eq:6.6}), as well as the fact that from Eq. (\ref{eq:1.1}) one has 
\begin{equation} \label{eq:6.14}
[L_0,L_+]= L_+\Rightarrow L_0 L_+ = L_+(L_0+1) \Rightarrow f(L_0) L_+
= L_+ f(L_0+1),
\end{equation}
\begin{equation} \label{eq:6.15}
[L_0, L_-]= -L_- \Rightarrow L_0 L_- = L_- (L_0-1) \Rightarrow f(L_0) L_-
= L_- f(L_0-1),
\end{equation}
where $f(x)$ is any function which can be written as a Taylor expansion 
in powers of $x$.  Indeed one has 
$$ [J_{+1},J_0]=-\frac{1}{\sqrt{[2]}}(q^{-L_0}L_{+}J_{0} -
J_{0}q^{-L_0}L_{+})=-\frac{1}{\sqrt{[2]}}
(L_{+}J_{0}-J_{0}L_{+})q^{-L_0-1} $$
\begin{equation} \label{eq:6.16}
=-\frac{1}{\sqrt{[2]}}\sqrt{[2]}J_{+1}q^{-L_0}q^{-L_0-1} =-q^{-2L_0+1}J_{+1},
\end{equation}
$$ [J_{-1},J_0]=\frac{1}{\sqrt{[2]}}(q^{-L_0}L_{-}J_{0} -
J_{0}q^{-L_0}L_{-})=\frac{1}{\sqrt{[2]}}
(L_{-}J_{0}-J_{0}L_{-})q^{-L_0+1} $$
\begin{equation} \label{eq:6.17}
=\frac{1}{\sqrt{[2]}}\sqrt{[2]}J_{-1}q^{-L_0}q^{-L_0+1}=q^{-2L_0-1}J_{-1}, 
\end{equation}
$$ [J_{+1},J_{-1}]=-\frac{1}{[2]}(q^{-L_0}L_{+}q^{-L_0}L_{-} -
q^{-L_0}L_{-}q^{-L_0}L_{+}) $$
$$ =-\frac{1}{[2]} (q^{-2L_0+1}L_{+}L_{-}-q^{-2L_0-1}L_{-}L_{+}) $$
\begin{equation} \label{eq:6.18}
=-\frac{1}{[2]}\,q^{-2L_0}(qL_{+}L_{-}-q^{-1}L_{-}L_{+}) =-q^{-2L_0}J_{0}, 
\end{equation}
or, in compact form 
\begin{equation} \label{eq:6.19} 
[J_{+1},J_0]=-q^{-2L_0+1}J_{+1}, \quad 
[J_{-1},J_0]=q^{-2L_0-1}J_{-1}, \quad 
[J_{+1},J_{-1}]=-q^{-2L_0}J_{0}. 
\end{equation}
In the limit $q\to 1$ these results reduce to the usual commutation relations 
related to spherical tensor operators under su(2)
\begin{equation} \label{eq:6.20}
[J_+, J_0]= - J_+, \quad [J_-, J_0]=J_-, \quad [J_+, J_-]=-J_0. 
\end{equation}
It is clear that the commutation relations of Eq. (\ref{eq:6.19}) are 
different from these of Eqs. (\ref{eq:1.1}), (\ref{eq:1.2}), as it is expected
since the commutation relations of Eq. (\ref{eq:6.20}) are different from 
the usual commutation relations of su(2), given in Eq. (\ref{eq:2.1}). 

One can now try to build out of these operators the scalar square of the 
angular momentum operator.  For this purpose one needs the definition 
of the tensor product of two irreducible tensor operators, which has the form
\cite{STK593,STK690,JPA6939,STK1068,STK1599,JPG1931}
\begin{equation} \label{eq:6.22}
[ A^{(q)}_{j_1} \otimes B^{(q)}_{j_2} ] ^{(1/q)}_{j,m} = \sum_{m_1,m_2} 
\langle j_1 m_1 j_2 m_2 | j m\rangle_{1/q} A^{(q)}_{j_1,m_1} B^{(q)}_{j_2,m_2}.
\end{equation}
One should observe that the irreducible tensor operators $A^{(q)}_{j_1}$ 
and $B^{(q)}_{j_2}$, which correspond to the deformation parameter $q$, 
 are combined into a new irreducible tensor operator
$[A^{(q)}_{j_1} \times B^{(q)}_{j_2}]^{(1/q)}_{j,m}$, 
which corresponds to the deformation parameter $1/q$,  through the use 
of the deformed Clebsch--Gordan coefficients 
$\langle j_1 m_1 j_2 m_2 | j m\rangle _{1/q}$,
which also correspond to the deformation parameter $1/q$. 

Analytic expressions for several $q$-deformed Clebsch--Gordan coefficients, 
as well as their symmetry proporties,  can be found in 
Refs. \cite{STK593,STK1068}. Using the general formulae of Refs. 
\cite{STK593,STK1068} 
we derive here the Clebsch--Gordan coefficients which we will immediately need 
\begin{equation} \label{eq:6.23}
\langle 11 10 | 11\rangle _q=q\sqrt{\frac{[2]}{[4]}}, \quad  
\langle 10 11 |11\rangle _q=-q^{-1}\sqrt{\frac{[2]}{[4]}}, 
\end{equation}
\begin{equation} \label{eq:6.24}
\langle 10 1-1 |1-1\rangle _q=q\sqrt{\frac{[2]}{[4]}} ,\quad
\langle 1-1 10 |1-1\rangle _q=-q^{-1}\sqrt{\frac{[2]}{[4]}},
\end{equation} 
\begin{equation} \label{eq:6.25} 
\langle 11 1-1 |10\rangle _q=\sqrt{\frac{[2]}{[4]}}, \quad
\langle 1-1 11 |10\rangle _q=-\sqrt{\frac{[2]}{[4]}}, \quad 
\langle 10 10 |10\rangle _q=(q-q^{-1})\sqrt{\frac{[2]}{[4]}}.
\end{equation}

Using the definition of Eq. (\ref{eq:6.22}), the Clebsch--Gordan coefficients
just given, as well as the commutation relations of Eq. (\ref{eq:6.19}), 
one finds the tensor products 
$$ [J\otimes J]_{1,+1}^{(1/q)}=
\langle 11 10 |11\rangle _{1/q} J_{+1}J_{0}+
\langle 10 11 |11\rangle _{1/q} J_{0}J_{+1}
=\sqrt{\frac{[2]}{[4]}}
\left\{q^{-1}J_{+1}J_{0}-qJ_{0}J_{+1}\right\} $$
$$ =\sqrt{\frac{[2]}{[4]}}
\left\{q^{-1}(J_{0}J_{+1}-q^{-2L_0+1}J_{+1})-qJ_{0}J_{+1}\right\}
=\sqrt{\frac{[2]}{[4]}}
\left\{(q^{-1}-q)J_0-q^{-2L_0}\right\}J_{+1} $$
\begin{equation} \label{eq:6.26}
=-\sqrt{\frac{[2]}{[4]}}
\left\{q^{-2L_0}+(q-q^{-1})J_0\right\}J_{+1}, 
\end{equation}
$$ [J\otimes J]_{1,-1}^{(1/q)}=
\langle 10 1-1 |1-1\rangle _{1/q}J_{0}J_{-1}+
\langle 1-1 10|1-1\rangle _{1/q} J_{-1}J_{0} $$
$$ =\sqrt{\frac{[2]}{[4]}}
\left\{q^{-1}J_{0}J_{-1}-qJ_{-1}J_{0}\right\} 
 =\sqrt{\frac{[2]}{[4]}}
\left\{q^{-1}J_{0}J_{-1}-q(J_{0}J_{-1}+q^{-2L_0-1}J_{-1})\right\}$$
\begin{equation} \label{eq:6.27} 
=\sqrt{\frac{[2]}{[4]}}
\left\{(q^{-1}-q)J_0-q^{-2L_0}\right\}J_{-1} 
=-\sqrt{\frac{[2]}{[4]}}
\left\{q^{-2L_0}+(q-q^{-1})J_0\right\}J_{-1}, 
\end{equation}
$$ [J\otimes J]_{1,0}^{(1/q)}=
\langle 11 1-1|10\rangle _{1/q}J_{+1}J_{-1}+
\langle 1-1 11|10\rangle _{1/q}J_{-1}J_{+1}+
\langle 10 10|10\rangle _{1/q}(J_{0})^2 $$
$$=\sqrt{\frac{[2]}{[4]}}\left\{J_{+1}J_{-1}-J_{-1}J_{+1}
+(q^{-1}-q)(J_{0})^2\right\}
=\sqrt{\frac{[2]}{[4]}}
\left\{-q^{-2L_0}J_{0}-(q-q^{-1})(J_{0})^2\right\} $$
\begin{equation} \label{eq:6.28}
=-\sqrt{\frac{[2]}{[4]}}
\left\{q^{-2L_0}+(q-q^{-1})J_0\right\}J_{0}. 
\end{equation}
We remark that all these tensor products are of the general form
\begin{equation} \label{eq:6.29} 
[J\otimes J]_{1,m}^{(1/q)}=
-\sqrt{\frac{[2]}{[4]}}\left\{q^{-2L_0}+(q-q^{-1})J_0\right\}J_{m}
= -\sqrt{\frac{[2]}{[4]}} Z J_m, 
\qquad\qquad m=0,\pm1
\end{equation}
where by definition 
\begin{equation} \label{eq:6.30}
Z=q^{-2L_0}+(q-q^{-1})J_0.
\end{equation}
One can now prove that the operator $Z$ is a scalar quantity, since it is 
a function of the second order Casimir operator of su$_q$(2), given in 
Eq. (\ref{eq:1.11}). Indeed one has 
$$ Z=q^{-2L_0}+(q-q^{-1})J_0
=q^{-2L_0}+\frac{(q-q^{-1})}{[2]}
\left\{q[2L_0]+(q-q^{-1})
\left(C_2^{(q)}-[L_0][L_0+1]\right)\right\} $$
$$=q^{-2L_0}+\frac{1}{[2]}\left\{q(q^{2L_0}-q^{-2L_0})
+(q-q^{-1})^2C_2^{(q)}-(q^{L_0}-q^{-L_0})(q^{L_0+1}-q^{-L_0-1})\right\}$$
$$=q^{-2L_0}+\frac{1}{[2]}\left\{\underline{q^{2L_0+1}}-q^{-2L_0+1} 
-\underline{
q^{2L_0+1}}+q+q^{-1}-q^{-2L_0-1}+(q-q^{-1})^2C_2^{(q)}\right\} $$
$$=q^{-2L_0}+\frac{1}{[2]}\left\{-q^{-2L_0}(q+q^{-1})
+(q+q^{-1})+(q-q^{-1})^2C_2^{(q)}\right\}$$
\begin{equation} \label{eq:6.31}
=q^{-2L_0}-q^{-2L_0}+1+\frac{(q-q^{-1})^2}{[2]}\,C_2^{(q)}
=1+\frac{(q-q^{-1})^2}{[2]}\,C_2^{(q)}
\end{equation}
or, in more compact form, 
\begin{equation} \label{eq:6.32} 
Z=q^{-2L_0}+(q-q^{-1})J_0=1+\frac{(q-q^{-1})^2}{[2]}\,C_2 ^{(q)}.
\end{equation}

Since $Z$ is a scalar quantity, symmetric under the exchange 
$q\leftrightarrow q^{-1}$ (as one can see from the last expression 
appearing in the last equation), Eq. (\ref{eq:6.29})  can be written 
in the form 
\begin{equation} \label{eq:6.33}
\left[ {J\over Z} \otimes {J\over Z} \right]_{1,m}^{(1/q)} = -\sqrt{[2]\over
 [4]} {J_m \over Z} \Rightarrow [J' \otimes J']_{1,m}^{(1/q)} = 
-\sqrt{[2]\over [4]} J'_m,
\end{equation}
where by definition
\begin{equation} \label{eq:6.34}
J'_m = {J_m \over Z}, \qquad m=+1,0,-1. 
\end{equation}
It is clear that the operators $J'_m$ also form 
an irreducible tensor operator, since $Z$ is a function of the second order 
Casimir $C_2^{(q)}$ of su$_q$(2), which commutes with the generators $L_+$, 
$L_-$, $L_0$ of su$_q$(2), and therefore does not affect the commutation
relations of Eqs. (\ref{eq:6.5}), (\ref{eq:6.6}). 

The scalar product of two irreducible tensor operators is defined as 
\cite{STK690,JPG1931} 
\begin{equation} \label{eq:6.35}
( A^{(q)}_j \cdot B^{(q)}_j )^{(1/q)} = (-1)^{-j} \sqrt{[2j+1]}
[ A^{(q)}_j \times B^{(q)}_j ]^{(1/q)}_{0,0} 
= \sum_m (-q)^{-m} A^{(q)}_{j,m} B^{(q)}_{j,-m}.
\end{equation}
Substituting the irreducible tensor operators $J_m$ in this definition 
we obtain \cite{STK690}
\begin{equation} \label{eq:6.36} 
(J \cdot J)^{(1/q)}= - \sqrt{[3]} [J\times J]_{0,0}^{(1/q)} =
{2\over [2]} C_2^{(q)} + {(q-q^{-1})^2 \over [2]^2} (C_2^{(q)})^2 
= {Z^2 -1 \over (q-q^{-1})^2},  
\end{equation}
where in the last step the identity 
\begin{equation} \label{eq:6.37}
Z^2-1 = (Z-1)(Z+1) = {(q-q^{-1})^2 \over [2]} C_2^{(q)} 
\left( 2+{(q-q^{-1})^2 \over [2]} C_2^{(q)} \right),
\end{equation}
has been used, obtained through use of Eq. (\ref{eq:6.32}). 
In the same way the irreducible tensor operators $J'_m$ give the result
\begin{equation} \label{eq:6.38}
\left( J' \cdot J' \right)^{(1/q)} = {1-Z^{-2} \over (q-q^{-1})^2} . 
\end{equation}
We have therefore determined the scalar square of the angular momentum 
operator. We can assume at this point that this quantity can be used 
(up to an overall constant) as the Hamiltonian for the description of 
rotational spectra, defining 
\begin{equation} \label{eq:6.39}
H = A {1-Z^{-2} \over (q-q^{-1})^2},
\end{equation}
where $A$ is a constant, which we also write in the form
\begin{equation} \label{eq:6.40} 
A={\hbar^2 \over 2 {\cal J}_0}
\end{equation} 
for future reference. 

The eigenvalues $\langle Z\rangle $ of the operator $Z$ 
in the basis $|\ell,m\rangle $ can be 
easily found from the last expression given in Eq. (\ref{eq:6.32}), 
using the eigenvalues of the Casimir operator $C_2^{(q)}$ in this basis, 
which are $[\ell] [\ell+1]$, as already mentioned in Sec. 2
\begin{equation} \label{eq:6.41} 
\langle Z\rangle = 1 + {(q-q^{-1})^2 \over [2]} [\ell] [\ell+1] = 
{1\over [2]}
(q^{2\ell+1}+q^{-2\ell-1}) = {1\over [2]} ([2\ell+2]-[2\ell]).
\end{equation}
The eigenvalues $\langle (J\cdot J)^{(1/q)}\rangle $ of the scalar quantity 
$(J \cdot J)^{(1/q)}$ can be 
found in a similar manner from Eq. (\ref{eq:6.36})
\begin{equation} \label{eq:6.42}
\langle (J \cdot J)^{(1/q)}\rangle  
= {2\over [2]} [\ell][\ell+1] + {(q-q^{-1})^2 \over [2]^2}
[\ell]^2 [\ell+1]^2 = {[2\ell] [2\ell+2] \over [2]^2} =
[\ell]_{q^2} [\ell+1]_{q^2} ,
\end{equation}
where by definition
\begin{equation} \label{eq:6.43} 
[x]_{q^2} = {q^{2x}-q^{-2x}\over q^2 -q^{-2}}.
\end{equation}
Finally, the eigenvalues $\langle H\rangle $ 
of the Hamiltonian can be found by substituting 
the eigenvalues of $Z$ from Eq. (\ref{eq:6.41}) into Eq. (\ref{eq:6.39})
$$ E = \langle H\rangle = A {1\over (q-q^{-1})^2} \left( 1-{[2]^2 \over 
(q^{2\ell+1}+q^{-2\ell-1})^2 } \right) $$ 
\begin{equation} \label{eq:6.44}
=A {1\over 4\sinh^2\tau } \left( 1-{\cosh^2\tau \over \cosh^2((2\ell+1)\tau)}
\right),   \qquad q=e^\tau , 
\end{equation}
where in the last step the identities 
\begin{equation} \label{eq:6.45} 
q-q^{-1}= 2\sinh\tau, \qquad [2] = q+q^{-1} = 2\cosh\tau,
\end{equation}
\begin{equation} \label{eq:6.46} 
q^{2\ell+1}+q^{-2\ell-1} =2\cosh ((2\ell+1)\tau),
\end{equation}
which are valid in the present case of $q=e^\tau$ with $\tau$ being real, 
have been used. 
In the same way one sees that 
\begin{equation} \label{eq:6.47}
\langle Z\rangle = {\cosh((2\ell+1)\tau)\over \cosh\tau}.
\end{equation}

The following comments are now in place: 

a) The last expression in Eq. (\ref{eq:6.42}) indicates that the eigenvalues 
of the scalar quantity $(J\cdot J)^{(1/q)}$ 
are equivalent to the eigenvalues of the Casimir operator of su$_q$(2) 
(which are $[\ell][\ell+1]$), up to a change 
in the deformation parameter from $q$ to $q^2$. 

b) From Eq. (\ref{eq:6.41}) it is clear that the eigenvalues of the 
scalar operator $Z$ go to the limiting value 1 as $q\rightarrow 1$. 
Therefore one can think of $Z$ as a ``unity'' operator. Furthermore the last 
expression in Eq. (\ref{eq:6.41}) indicates that 
$\langle Z\rangle $ is behaving like 
a ``measure'' of the unit of angular momentum in the deformed case. 

\section{Rotational invariance of the su$_q$(2) ITO Hamiltonian} 

In this section the notation and tools of Sec. 3 will be used once more. 
We wish to prove that the Hamiltonian of Eq. (\ref{eq:6.39}) commutes 
with the generators $\hat l_+$, $\hat l_-$, $\hat l_0$ of the usual su(2) 
algebra, i.e.
with the usual angular momentum operators. 
Taking into account Eq. (\ref{eq:6.32}) we see that 
acting on the ``classical'' basis described in Sec. 3 we have
\begin{equation} \label{eq:8.1} 
\hat Z |l, {\rm m}\rangle _c = \left( 1+ {(q-q^{-1})^2 \over [2]} \hat 
C_2^{(q)}\right)
| l, {\rm m}\rangle _c = \left(1+{(q-q^{-1})^2 \over [2]}[l] [l+1]\right)
|l, {\rm m}\rangle _c.
\end{equation}
Then using Eq. (\ref{eq:6.39}) we see that 
$$ \hat H |l, {\rm m}\rangle _c = {A\over (q-q^{-1})^2} \left(1-{1\over \hat 
Z^2}\right) 
|l, {\rm m}\rangle _c $$
\begin{equation} \label{eq:8.2} 
= {A\over (q-q^{-1})^2} \left(1-{1\over 
\left( 1+ {(q-q^{-1})^2 \over [2]} [l] [l+1]\right)^2} \right) 
|l, {\rm m}\rangle _c.
\end{equation}
Using this result, as well as Eq. (\ref{eq:2.4}), one finds 
$$[\hat H, \hat l_+] |l, {\rm m}\rangle _c = \hat H \hat l_+ |l, 
{\rm m}\rangle _c -\hat l_+ \hat H |l, {\rm m}\rangle _c $$
$$= \hat H \sqrt{(l-m)(l+m+1)} |l, {\rm m}+1\rangle _c $$
$$- \hat l_+ {A\over (q-q^{-1})^2} 
\left(1 -{1\over \left(1+{(q-q^{-1})^2\over [2]} [l][l+1] \right)^2}\right) 
|l, {\rm m}\rangle _c$$
$$={A\over (q-q^{-1})^2} \left(1-{1\over \left(1+{(q-q^{-1})^2 \over [2]} 
[l][l+1]\right)^2}\right) \sqrt{(l-m)(l+m+1)} |l, {\rm m}+1\rangle _c $$
\begin{equation} \label{eq:8.3} 
-\sqrt{(l-m)(l+m+1)} {A\over (q-q^{-1})^2} \left(1-{1\over \left(1+ 
{(q-q^{-1})^2\over [2]} [l][l+1]\right)^2 }\right) |l, {\rm m}+1\rangle_c =0. 
\end{equation}
In exactly the same way, using Eqs. (\ref{eq:2.4}) and (\ref{eq:8.2}), 
one finds that 
\begin{equation} \label{eq:8.4} 
[\hat H, \hat l_-] |l, {\rm m}\rangle_c =0, \qquad [\hat H, \hat l_0] 
|l, {\rm m}\rangle_c =0.
\end{equation}
We have thus proved that the Hamiltonian of Eq. (\ref{eq:6.39}) is invariant 
under usual angular momentum. This result is expected, since the Hamiltonian 
is a function of the operator $\hat Z$, which in turn (as seen from Eq. 
(\ref{eq:6.32})~) is a function of 
the second order Casimir operator of su$_q$(2), $\hat C_2^{(q)}$, which was 
proved to be rotationally invariant in Section 3. 

Since the Hamiltonian of Eq. (\ref{eq:6.39}) is rotationally invariant, 
it should be possible to express it as a function of $\hat C_2$ 
(the second order Casimir 
operator of su(2)~). It should also be possible to express the eigenvalues 
of the Hamiltonian of Eq. (\ref{eq:6.39}) as a function of $l(l+1)$, i.e. 
as a function of the eigenvalues of $\hat C_2$. This task will be undertaken 
in the following section. 

For completeness we mention that using Eqs. (\ref{eq:2.17}) and 
(\ref{eq:8.2}) one can prove in an analogous way that 
\begin{equation} \label{eq:8.5}
[\hat H, \hat L_+] |l, {\rm m}\rangle_c =0, \qquad 
[\hat H, \hat L_-] |l, {\rm m}\rangle_c =0, \qquad 
[\hat H, \hat L_0] |l, {\rm m}\rangle_c =0,
\end{equation}
i.e. that the Hamiltonian of Eq. (\ref{eq:6.39}) commutes with the 
generators of su$_q$(2) as well. Then from Eq. (\ref{eq:6.12}) it is 
clear that in addition one has 
\begin{equation} \label{eq:8.6} 
[\hat H, \hat J_+] |l, {\rm m}\rangle _c=0, \qquad 
[\hat H, \hat J_-] |l, {\rm m}\rangle_c =0, \qquad  
[\hat H, \hat J_0] |l, {\rm m}\rangle _c =0. 
\end{equation}
Then from Eqs. (\ref{eq:6.34}) and (\ref{eq:8.1}) one furthermore obtains
\begin{equation} \label{eq:8.7}
[\hat H, \hat J'_+]|l, {\rm m}\rangle _c =0, \qquad 
[\hat H, \hat J'_-] |l, {\rm m}\rangle _c =0, \qquad 
[\hat H, \hat J'_0] |l, {\rm m}\rangle _c=0. 
\end{equation}

\section{Exact expansion of the su$_q$(2) ITO spectrum}

Since the Hamiltonian of Eq. (\ref{eq:6.39}) is invariant under su(2), 
as we have seen in the last section, it should be possible to write 
its eigenvalues (given in Eq. (\ref{eq:6.44})~) as an expansion in terms 
of $\ell(\ell+1)$. This is a nontrivial task, since in Eq. (\ref{eq:6.44})
a function of the variable $\ell$ appears, while we are in need of a function
of the variable $\ell(\ell+1)$, which is related to the length of the 
angular momentum vector. 
For this purpose it turns out that one should use 
the Taylor expansion \cite{AbrSte}
\begin{equation}\label{eq:9.1}
\tanh x=\sum_{n=1}^{\infty}\frac{2^{2n}(2^{2n}-1)B_{2n}}{(2n)!}
\,x^{2n-1}
=\sum_{n=0}^{\infty}\frac{2^{2n+2}(2^{2n+2}-1)B_{2n+2}}{(2n+2)!}
\,x^{2n+1}
\quad,\qquad |x|<\frac{\pi}{2}, 
\end{equation}
where $B_{n}$ are the Bernoulli numbers \cite{AbrSte}, defined through the
generating function
\begin{equation} \label{eq:9.2} 
\frac{x}{e^x-1}=\sum_{n=0}^{\infty}B_n\frac{x^n}{n!}, 
\end{equation}
the first few of them being 
$$ B_0=1,\quad B_1=-\frac{1}{2},\quad B_2=\frac{1}{6},
\quad B_4=-\frac{1}{30},\quad B_6=\frac{1}{42},
\quad B_8=-\frac{1}{30},\quad B_{10}=\frac{5}{66},\ldots, $$
\begin{equation} \label{eq:9.3}
B_{2n+1}=0 \quad {\rm for} \quad n=1,2,\ldots
\end{equation}
From Eq. (\ref{eq:9.1}) the following identities, concerning the derivatives 
of $\tanh x$, occur 
\begin{equation} \label{eq:9.4}
(\tanh x)'=\frac{1}{\cosh ^2x}=1-\tanh ^2x
=\sum_{n=0}^{\infty}
\frac{2^{2n+2}(2^{2n+2}-1)B_{2n+2}}{(2n)!(2n+2)}\,x^{2n}, 
\end{equation}
\begin{equation} \label{eq:9.5}
(\tanh x)''= -2{\tanh x \over \cosh^2x} = -2 {\sinh x \over \cosh^2 x} 
=\sum_{n=0}^{\infty}
\frac{2^{2n+4}(2^{2n+4}-1)B_{2n+4}}{(2n+1)!(2n+4)}\,x^{2n+1}.
\end{equation}
From these equations the following auxiliary identities occur 
\begin{equation} \label{eq:9.6}
\frac{\sinh x}{x \ \cosh^3x}=-\frac{1}{2x}(\tanh x)''
=\sum_{n=0}^{\infty}
\frac{2^{2n+3}(1-2^{2n+4})B_{2n+4}}{(2n+1)!(2n+4)}\,x^{2n}, 
\end{equation}
\begin{equation} \label{eq:9.7}
\tanh^2x=1-\frac{1}{\cosh^2x}
=\sum_{n=0}^{\infty}
\frac{2^{2n+4}(1-2^{2n+4})B_{2n+4}}{(2n+2)!(2n+4)}\,x^{2n+2}. 
\end{equation}
The expression for the energy, given in Eq. (\ref{eq:6.44}), can be put 
in the form  
\begin{equation} \label{eq:9.8}
{E\over A} =
\left(\frac{{\rm cosh}^2\tau\,.\tau^2}{{\rm sinh}^2\tau}\right)
\frac{1}{(2\tau)^2}\left\{\frac{1}{{\rm cosh}^2\tau}
-\frac{1}{{\rm cosh}^2((2\ell+1)\tau)}\right\}. 
\end{equation}
Denoting
\begin{equation} \label{eq:9.9}
z=(2\ell+1)\tau, \qquad x=\ell(\ell+1), 
\end{equation}
which imply 
\begin{equation} \label{eq:9.10} 
z^2=(4x+1)\tau^2, \qquad\qquad 
z^{2n}=\tau^{2n}\sum_{k=0}^{n}{n\choose k}\,2^{2k}\,x^k, 
\end{equation}
(the latter through use of the standard binomial formula), 
one obtains from Eq. (\ref{eq:9.4}) the expansion 
$$ \frac{1}{\cosh^2((2\ell+1)\tau)}=
\frac{1}{\cosh^2z}=\sum_{n=0}^{\infty}
\frac{2^{2n+2}(2^{2n+2}-1)B_{2n+2}}{(2n)!(2n+2)}\,z^{2n} $$
\begin{equation} \label{eq:9.11} 
=\sum_{n=0}^{\infty}\;
\underbrace{\frac{2^{2n+2}(2^{2n+2}-1)B_{2n+2}}{(2n)!(2n+2)}
\,\tau^{2n}}_{a_n}
\;\sum_{k=0}^{n}\;
\underbrace{{n\choose k}\,2^{2k}}_{b_{n,k}}\;x^k
=\sum_{n=0}^{\infty} a_n\, \sum_{k=0}^n \, b_{n,k}\,  x^k. 
\end{equation}
The double sum appearing in the last expression can be rearranged using the 
general procedure 
$$ S=\sum_{n=0}^{\infty}a_n\sum_{k=0}^n b_{n,k}\,x^k $$
$$ = a_0b_{00}+a_1(b_{10}+b_{11}x)+a_2(b_{20}+b_{21}x+b_{22}x^2)
+a_3(b_{30}+b_{31}x+b_{32}x^2+b_{33}x^3)+\ldots $$
$$=(a_0b_{00}+a_1b_{10}+a_2b_{20}+a_3b_{30}+\ldots)
+(a_1b_{11}+a_2b_{21}+a_3b_{31}+a_4b_{41}+\ldots)\,x $$
$$+(a_2b_{22}+a_3b_{32}+a_4b_{42}+a_5b_{52}+\ldots)\,x^2
+(a_3b_{33}+a_4b_{43}+a_5b_{53}+a_6b_{63}+\ldots)\,x^3
\,+\,\ldots $$
\begin{equation}\label{eq:9.12}
=\sum_{n=0}^{\infty}\,\underbrace{\left\{
\sum_{k=n}^{\infty} a_{k}\,b_{k,n}\right\}}_{c_n}\,x^n
=\sum_{n=0}^{\infty}c_{n}\,x^n, 
\end{equation}
where
\begin{equation} \label{eq:9.13} 
c_{n}=\sum_{k=n}^{\infty} a_{k}\,b_{k,n}
=\sum_{k=0}^{\infty} a_{n+k}\,b_{n+k,n}. 
\end{equation}
Applying this general procedure in the case of Eq. (\ref{eq:9.11}) we obtain 
\begin{equation} \label{eq:9.14} 
{1\over \cosh^2((2\ell+1)\tau)} = {1\over \cosh^2z} = \sum_{n=0}^\infty 
c_n x^n, 
\end{equation}
where
$$ c_{n}=\sum_{k=0}^{\infty} a_{n+k}\,b_{n+k,n}\nonumber\\[2ex]
=\sum_{k=0}^{\infty}
\frac{2^{2n+2k+2}(2^{2n+2k+2}-1)B_{2n+2k+2}}{(2n+2k)!(2n+2k+2)}
\,\tau^{2n+2k}\,{n+k\choose n}\,2^{2n} $$
\begin{equation} \label{eq:9.15} 
=(2\tau)^{2n}\,\sum_{k=0}^{\infty}
\frac{2^{2n+2k+2}(2^{2n+2k+2}-1)B_{2n+2k+2}}{(2n+2k)!(2n+2k+2)}
\,{n+k\choose n}\,\tau^{2k}. 
\end{equation}
The first term in Eq. (\ref{eq:9.14}) is
\begin{equation} \label{eq:9.16} 
c_0=\sum_{k=0}^{\infty}
\frac{2^{2k+2}(2^{2k+2}-1)B_{2k+2}}{(2k)!(2k+2)}
\,\tau^{2k}=\frac{1}{{\rm cosh}^2\tau}.
\end{equation}
Then one has 
$$ \frac{1}{(2\tau)^2}\left\{\frac{1}{{\rm cosh}^2\tau}
-\frac{1}{{\rm cosh}^2((2\ell+1)\tau)}\right\}
=-\frac{1}{(2\tau)^2}\sum_{n=1}^\infty c_n\, x^n $$ 
\begin{equation} \label{eq:9.17} 
=-\frac{1}{(2\tau)^2}\sum_{n=0}^{\infty}c_{n+1}\,x^{n+1}
=\sum_{n=0}^{\infty}d_{n}\,x^{n+1},
\end{equation}
where the coefficients $d_n$ are 
\begin{equation}  \label{eq:9.18} 
d_{n}=-\frac{1}{(2\tau)^2}\,c_{n+1}
=\frac{(-1)^n(2\tau)^n}{(n+1)!}\;f_{n}(\tau), 
\qquad\qquad n=0,1,2,\ldots
\end{equation}
with
\begin{equation} \label{eq:9.19} 
f_{n}(\tau)=(-1)^{n+1}(2\tau)^n(n+1)!
\sum_{k=0}^{\infty}
\frac{2^{2n+2k+4}(2^{2n+2k+4}-1)B_{2n+2k+4}}{(2n+2k+2)!(2n+2k+4)}
\,{n+k+1\choose n+1}\,\tau^{2k}. 
\end{equation}
For $n=0$ one has 
\begin{equation} \label{eq:9.20} 
f_{0}(\tau)=-\sum_{k=0}^{\infty}
\frac{2^{2k+4}(2^{2k+4}-1)B_{2k+4}}{(2k+2)!(2k+4)}
\,(k+1)\,\tau^{2k}
=\frac{\sinh\tau}{\tau\cosh^3\tau},
\end{equation}
where in the last step Eq. (\ref{eq:9.6}) has been used. 
It is worth noticing that 
\begin{equation} \label{eq:9.21} 
f_n(\tau)= (-1)^n \tau^n \left( {1\over \tau} {d \over d\tau}\right)^n 
f_0(\tau). 
\end{equation}
With the help of Eqs. (\ref{eq:9.17}) and (\ref{eq:9.18}), 
the spectrum of Eq. (\ref{eq:9.8}) is put into the form 
\begin{equation} \label{eq:9.22} 
{E\over A}
=\left(\frac{\tau^2 \, \cosh^2\tau}{\sinh^2\tau}\right)
\sum_{n=0}^{\infty}
\frac{(-1)^n(2\tau)^n}{(n+1)!}\;f_{n}(\tau)
\,(\ell(\ell+1))^{n+1},
\end{equation}
since $x=\ell(\ell+1)$ from Eq. (\ref{eq:9.9}). 
It is clear that Eq. (\ref{eq:9.22}) is an expansion in terms of 
$\ell(\ell+1)$, as expected. 

\section{Approximate expansion of the su$_q$(2) ITO spectrum} 

In the limit of $|\tau|<<1$ one is entitled to keep in Eq. (\ref{eq:9.19}) 
only the term with $k=0$. Then the function $f_n(\tau)$ takes the form 
\begin{equation} \label{eq:10.1} 
f_n(\tau) \rightarrow {(-1)^{n+1} 2^{2n+2} (2^{2n+4}-1)
B_{2n+4} \over (2n+1)!!  (n+2)}  \tau^n,
\end{equation}
where the Bernoulli numbers appear again and use of the identity 
\begin{equation} \label{eq:10.2} 
(2n+2)! = 2^{n+1} (n+1)! (2n+1)!!
\end{equation} 
has been made. 
Taking into account the Taylor expansions 
\begin{equation} \label{eq:10.3} 
\sinh x = x + {x^3\over 3!} + {x^5 \over 5!} + \cdots, \qquad
\cosh x = 1 + {x^2 \over 2!} + {x^4 \over 4!} + \cdots,
\end{equation}
and keeping only the lowest order terms, one easily sees that 
Eq. (\ref{eq:9.22}) is put in the form 
\begin{equation} \label{eq:10.4} 
{E\over A} \approx \sum_{n=0}^\infty  {2^{2n+4} 
(1-2^{2n+4}) B_{2n+4} \over (2n+2)! (2n+4)} (2\tau)^{2n}  (\ell(\ell+1))^{k+1},
\end{equation}
where use of the identity of Eq. (\ref{eq:10.2}) has been made once more
and use of the fact that 
\begin{equation} \label{eq:10.5}
{\tau^2 \cosh^2\tau \over \sinh^2\tau} \approx 1 
\quad {\rm for} \quad |\tau|<<1
\end{equation}
has been made.  
Comparing this result with Eq. (\ref{eq:9.7})
and making the identifications
\begin{equation} \label{eq:10.6} 
x=2\tau \sqrt{\ell (\ell+1)} = 2\tau \xi, \qquad \xi=\sqrt{\ell(\ell+1)},
\end{equation}
Eq. (\ref{eq:10.4}) is put into the compact form
\begin{equation} \label{eq:10.7} 
E\approx {A\over (2\tau)^2} \tanh^2(2\tau\sqrt{\ell(\ell+1)})
= {A\over (2\tau)^2} \tanh^2(2\tau\xi), \qquad q=e^\tau. 
\end{equation}
The extended form of the Taylor expansion of $E$ is easily obtained from 
Eq. (\ref{eq:10.4})
\begin{equation} \label{eq:10.8} 
E\approx A \left( \ell(\ell+1)-{2\over 3} (2\tau)^2 (\ell(\ell+1))^2 
+{17\over 45} (2\tau)^4 (\ell(\ell+1))^3 -{62\over 315} (2\tau)^6 
(\ell(\ell+1))^4 +\cdots\right). 
\end{equation}
Eq. (\ref{eq:10.7}) will be referred to as the ``hyperbolic tangent formula''. 

\section{Analytic expressions based on the approximate expansion
of the su$_q$(2) ITO spectrum}

We are now going to derive analytic formulae for the rotational frequency 
$\omega$ and the kinematic moment of inertia ${\cal J}$, based on the 
approximate expression for the energy given in Eq. (\ref{eq:10.7}).  
From Eqs. (\ref{eq:5.1}) and (\ref{eq:5.2}) 
one immediately obtains 
\begin{equation} \label{eq:11.1} 
\hbar \omega = {\partial E\over \partial \xi} = {A\over \tau} {\sinh(2\tau\xi)
\over \cosh^3(2\tau\xi)} = {A\over \tau} \tanh(2\tau \xi) 
(1-\tanh^2(2\tau\xi)), 
\end{equation}
\begin{equation} \label{eq:11.2} 
{\hbar^2 \over 2 {\cal J}} = {\partial E\over \partial \eta} ={1\over 2\xi}
{\partial E\over \partial \xi} = {A \over 2\tau\xi} {\sinh(2\tau\xi) \over
\cosh^3(2\tau\xi)} ={A\over 2\tau\xi} \tanh(2\tau\xi) (1-\tanh^2(2\tau\xi)),
\end{equation}
where by definition $\eta=\ell(\ell+1)=\xi^2$, as in Eq. (\ref{eq:5.3}).
Using the expressions for $E$ and $\omega$ given in Eqs. (\ref{eq:10.7}) 
and (\ref{eq:11.1}) one 
can easily verify that 
\begin{equation}\label{eq:11.3} 
{ {\cal J}_0 \omega^2 \over 2} = E \left( 1-{(2\tau)^2 \over A}E\right)^2,
\end{equation}
where use of Eq. (\ref{eq:1.15})
and of the identities 
\begin{equation} \label{eq:11.4}
\cosh^2x-\sinh^2x=1, \qquad {1\over \cosh^2 x}= 1-\tanh^2 x, 
\end{equation} 
 has also been made. 
Defining 
\begin{equation} \label{eq:11.5}
\varepsilon= {(2\tau)^2\over A} E = \tanh^2(2\tau\xi),
\end{equation}
\begin{equation} \label{eq:11.6} 
t={\hbar \tau\over A} = {2 {\cal J}_0 \over \hbar} \tau, 
\end{equation}
where $t$ is a constant having dimensions of time, 
Eq. (\ref{eq:11.3}) takes the form
\begin{equation} \label{eq:11.7}
(\omega t)^2 = \varepsilon (1-\varepsilon)^2 = \varepsilon -2\varepsilon^2 
+\varepsilon^3.
\end{equation}
From this equation one can determine $\varepsilon$ as a function of $\omega t$,
in the following way. One can define 
\begin{equation} \label{eq:11.16}
s(x)=1-\varepsilon(x) \Rightarrow \varepsilon(x)=1-s(x), 
\qquad  x=(\omega t)^2.
\end{equation} 
Then Eq. (\ref{eq:11.7}) takes the form 
\begin{equation} \label{eq:11.17}
\varepsilon(1-\varepsilon)^2 = (1-s) s^2 = x \Rightarrow  s^2 -s^3 =x. 
\end{equation}
From Eq. (\ref{eq:11.7}) it is clear that 
\begin{equation} \label{eq:11.18}
\varepsilon(\omega =0)=0, 
\end{equation}
which immediately implies 
\begin{equation} \label{eq:11.19} 
s(x=0) =1. 
\end{equation}
One can now try to express $s(x)$ as a power series in $x$, having the form
\begin{equation} \label{eq:11.20}
s(x)= 1 + a_1 x^2 + a_2 x^3 + a_3 x^4 + \ldots
\end{equation}
For a series of this form one can use the fact that 
$s^2(x)$ is of the form \cite{AbrSte}
\begin{equation} \label{eq:11.21} 
s^2(x)=1+b_1 x+b_2 x^2+b_3 x^3+\ldots,
\end{equation}
where the coefficients $b_n$ are given by the recursion relation
\begin{equation} \label{eq:11.22} 
b_n=\frac{1}{n}\sum_{k=1}^{n}(3k-n)a_k\,b_{n-k}\quad,\qquad
n\geq1, \qquad b_0=1, 
\end{equation}
as well as the fact that $s^3(x)$ is of the form \cite{AbrSte} 
\begin{equation} \label{eq:11.23} 
s^3(x)=1+c_1 x+c_2 x^2+c_3 x^3+\ldots,
\end{equation}
where the coefficients $c_n$ are given by the recursion relation
\begin{equation} \label{eq:11.24} 
c_n=\frac{1}{n}\sum_{k=1}^{n}(4k-n)a_k\,c_{n-k}\quad,\qquad
n\geq1, \qquad c_0=1,  
\end{equation}
the explicit form of the first few coefficints being
\begin{equation} \label{eq:11.25} 
b_1=2a_1, \qquad  c_1=3a_1,
\end{equation}
\begin{equation}\label{eq:11.26}
b_2=a_1^2+2a_2, \qquad c_2=3(a_1^2+a_2),
\end{equation}
\begin{equation}\label{eq:11.27}
b_3=2(a_1a_2+a_3),\qquad  c_3=a_1^3+6a_1a_2+3a_3,
\end{equation}
\begin{equation}\label{eq:11.28}
b_4=a_2^2+2a_1a_3+2a_4, \qquad 
c_4=3(a_1^2a_2+a_2^2+2a_1a_3+a_4),
\end{equation}
\begin{equation}\label{eq:11.29}
b_5=2(a_2a_3+a_1a_4+a_5),\qquad 
c_5=3(a_1a_2^2+a_1^2a_3+2a_2a_3+2a_1a_4+a_5).
\end{equation}

The coefficients in Eq. (\ref{eq:11.20}) can now be determined 
by considering Eq. (\ref{eq:11.17}) written in the form 
\begin{equation} \label{eq:11.30}
s^2(x)-s^3(x)=d_1 x+d_2 x^2+d_3 x^3+\ldots \equiv x, 
\end{equation}
which implies that 
\begin{equation} \label{eq:11.31} 
d_1=1 \qquad\mbox{and}\qquad
d_0=d_2=d_3=\ldots=0.
\end{equation}
The first few  coefficients in Eq. (\ref{eq:11.30}) are then
\begin{equation} \label{eq:11.32}
d_1=2a_1-3a_1=-a_1=1 \quad\Rightarrow\quad a_1=-1, 
\end{equation}
\begin{equation} \label{eq:11.33}
d_2=(a_1^2+2a_2)-3(a_1^2+a_2)=-2-a_2=0 \quad\Rightarrow\quad a_2=-2, 
\end{equation}
\begin{equation} \label{eq:11.34} 
d_3=2(a_1a_2+a_3)-(a_1^3+6a_1a_2+3a_3)=-7-a_3=0
\quad\Rightarrow\quad a_3=-7.
\end{equation}
By this procedure one obtains
\begin{equation} \label{eq:11.35}
s(x)=1-x-2x^2-7x^3-30x^4-143x^5-\ldots, 
\end{equation}
and
\begin{equation} \label{eq:11.36}
\varepsilon(x)=1-s(x)=x+2x^2+7x^3+30x^4+143x^5+\ldots, 
\end{equation}
which, using Eq. (\ref{eq:11.16}), takes the form 
\begin{equation} \label{eq:11.37}
\varepsilon(\omega)=(\omega t)^2+2(\omega t)^4+7(\omega t)^6
+30(\omega t)^8+143(\omega t)^{10}+\ldots
\end{equation}

It is clear that this expression corresponds to a real root of the cubic 
equation of Eq. (\ref{eq:11.7}), which is of the form 
\begin{equation} \label{eq:11.38}
\varepsilon^3 + f_2 \varepsilon^2 + f_1 \varepsilon + f_0 =0, 
\end{equation}
with 
\begin{equation} \label{eq:11.39}
f_2 = -2, \quad f_1 = 1, \quad f_0 = -(\omega t)^2. 
\end{equation}
Using the standard way of solving a cubic equation \cite{AbrSte} 
one has 
\begin{equation} \label{eq:11.40} 
g= {1\over 3} f_1 -{1\over 9} f_2^2 = -{1\over 9} ,
\end{equation}
\begin{equation} \label{eq:11.41} 
h={1\over 6} (f_1 f_2 -3 f_0) -{1\over 27} f_2^3= {1\over 2} (\omega t)^2 
-{1\over 9}, 
\end{equation}
while the discriminant is 
\begin{equation} \label{eq:11.42} 
D= g^3 + h^2 = \left({1\over 2} (\omega t)^2 -{1\over 9}\right)^2 
-\left({1\over 27}\right)^2 = \left({1\over 2} (\omega t)^2 -{2\over 27}
\right) \left({1\over 2}(\omega t)^2 -{4\over 27}\right). 
\end{equation}
One obtains three real roots when $D<0$ (i.e. when $4/27 \leq
 (\omega t)^2 \leq 8/27$, while for $D>0$ (i.e. for 
$(\omega t)^2> 8/27$ or for $(\omega t)^2 < 4/27$) one has only one 
real root. In the case of rotational spectra it is clear that we are 
interested in the region including $\omega =0$, i.e. the relevant region is 
$0 \leq (\omega t)^2 < 4/27$, in which only one real root exists. 
Using the standard procedure \cite{AbrSte} one can write in this case 
the explicit form of the real root, expand the square and cubic roots 
appearing there, and verify that the Taylor expansion of the root is 
of the form given in Eq. (\ref{eq:11.37}). 
 
Using Eqs. (\ref{eq:11.5}) and (\ref{eq:11.37}) one finally obtains 
the expansion of the energy in terms of powers of $\omega ^2$  
\begin{equation} \label{eq:11.8}
E={A\over (2\tau)^2} \varepsilon = {A\over (2\tau)^2} ( (\omega t)^2 
+2(\omega t)^4 +7 (\omega t)^6 + 30 (\omega t)^8 + 143 (\omega t)^{10}+
\ldots). 
\end{equation}  

On the other hand from Eq. (\ref{eq:11.2}) using Eq. (\ref{eq:1.15}) 
one obtains
\begin{equation} \label{eq:11.9} 
{ {\cal J} \over {\cal J}_0} = {2\tau \xi \over \tanh(2\tau\xi) 
(1-\tanh^2(2\tau\xi))} = 
{{\rm arctanh}(\sqrt{\varepsilon})\over 
\sqrt{\varepsilon}(1-\varepsilon)},
\end{equation}
where in the last step Eq. (\ref{eq:11.5}) has been taken into account. 
In the case of $0<\varepsilon<1$ (which guarantees that the Taylor 
expansion of ${\rm arctanh}(\sqrt{\varepsilon})$ is possible)
one can use the expansion \cite{AbrSte}
\begin{equation} \label{eq:11.43} 
{\rm arctanh} x = \sum_{n=0}^\infty {x^{2n+1} \over 2n+1}, \quad |x|<1.  
\end{equation}
In addition the following expansion holds 
\begin{equation} \label{eq:11.44} 
{1\over 1-x^2} = \sum_{n=0}^\infty x^{2n}, \quad |x|<1. 
\end{equation}
Using the general result \cite{AbrSte} that the series 
\begin{equation} \label{eq:11.45}
s_1(x) = 1 + a_1 x + a_2 x^2 + a_3 x^3 +\ldots
\end{equation}
and
\begin{equation} \label{eq:11.46} 
s_2(x)= 1 + b_1 x + b_2 x^2 + b_3 x^3 + \ldots 
\end{equation}
can be combined into 
\begin{equation} \label{eq:11.47} 
s_3(x) = s_1(x) s_2(x) = \sum_{n=0}^\infty c_n x^n
\end{equation}
with 
\begin{equation} \label{eq:11.48} 
c_n=\sum_{k=0}^n a_k b_{n-k}
\end{equation}
one obtains from Eqs. (\ref{eq:11.43}) and (\ref{eq:11.44})
\begin{equation} \label{eq:11.49} 
{ {\rm arctanh} x \over x} {1\over 1-x^2} = 
\sum_{n=0}^{\infty}\underbrace{\frac{1}{2n+1}}_{a_n}\,x^{2n}
\sum_{n=0}^{\infty}\underbrace{1}_{b_n}\,x^{2n}
=\sum_{n=0}^{\infty}c_n\,x^{2n}
\qquad,\quad x^2<1
\end{equation} 
with
\begin{equation} \label{eq:11.50} 
c_n=\sum_{k=0}^{n}a_k\,b_{n-k}=\sum_{k=0}^{n}\frac{1}{2k+1},
\end{equation}
the first few coefficients being 
\begin{equation} \label{eq:11.51} 
c_0=1,\quad c_1=\frac{4}{3},\quad c_2=\frac{23}{15},\quad
c_3=\frac{176}{105},\quad c_4=\frac{563}{315},\quad
c_5=\frac{6508}{3465}.
\end{equation}

Using Eq. (\ref{eq:11.49}) with $x=\varepsilon$ one can put Eq. 
(\ref{eq:11.9}) in the form 
\begin{equation} \label{eq:11.10} 
{ {\cal J} \over {\cal J}_0} = \sum_{n=0}^\infty c_n \varepsilon^n,
\end{equation}
where the coefficients are the ones given in Eqs. (\ref{eq:11.50}), 
(\ref{eq:11.51}). 

Eq. (\ref{eq:11.10}) is written analytically as 
\begin{equation} \label{eq:11.13} 
{ {\cal J} \over {\cal J}_0} = 1 +{4\over 3} \varepsilon +{23\over 15} 
\varepsilon^2 + {176\over 105} \varepsilon^3 + {563\over 315} \varepsilon^4 
+\ldots,
\end{equation}
which can be rewritten with the help of Eq. (\ref{eq:11.8}) as 
\begin{equation} \label{eq:11.14} 
{ {\cal J} \over {\cal J}_0} = 1 + {4\over 3} (\omega t)^2 + {21\over 5} 
(\omega t)^4 + {120\over 7} (\omega t)^6 + {715\over 9} (\omega t)^8 
+{4368\over 11} (\omega t)^{10} +\ldots
\end{equation} 
Using Eq. (\ref{eq:5.4}) one then additionally has 
$$\xi = \sqrt{\ell (\ell+1)} = { {\cal J} \omega \over \hbar} $$
\begin{equation} \label{eq:11.15} 
={ {\cal J}_0 \over \hbar} \omega \left( 1 + {4\over 3} (\omega t)^2 + 
{21\over 5} (\omega t)^4 + {120\over 7} (\omega t)^6 + {715\over 9} 
(\omega t)^8 + {4368\over 11} (\omega t)^{10}+ \ldots \right)
\end{equation} 
Eqs. (\ref{eq:11.8}) and (\ref{eq:11.15}) give the energy and the quantity 
$\sqrt{\ell(\ell+1)}$ 
as series in powers of the rotational frequency $\omega$, thus making 
contact between the present approach and the Harris formalism \cite{Harris}. 

\section{Numerical tests}

The formulae developed in the previous sections will be now tested against 
the experimental spectra of the Th isotopes 
\cite{Th222,Th224,Th226,Th228,Cocks}, which range from 
vibrational ($^{222}$Th with $R_4=E(4)/E(2)=2.399$) to clearly rotational 
($^{234}$Th with $R_4=3.308$). The purpose of this study is two-fold:

a) To test the quality of the approximations used in Secs. 5 and 10. 

b) To test the agreement between theoretical predictions 
and experimental data. 
 
The standard rotational expansion,
\begin{equation} \label{eq:12.1}
E= A \ell(\ell+1) + B (\ell(\ell+1))^2 + C(\ell(\ell+1))^3 
+ D (\ell(\ell+1))^4 +\ldots,
\end{equation}
from which only the first two terms will be included in order to keep 
the number of parameters equal to two, 
as well as the Holmberg--Lipas two-parameter expression \cite{Lipas}
\begin{equation} \label{eq:12.2}
E= a(\sqrt{1+b \ell(\ell+1)}-1),
\end{equation} 
which is known to give the best fits to experimental rotational nuclear
spectra among all two-parameter expressions \cite{Casten}, will be included 
in the test for comparison. 
For brevity we are going to use the following terminology: \hfill\break
Model I for Eq. (\ref{eq:1.17}) (original su$_q$(2) formula), \hfill\break
Model I$'$ for Eq. (\ref{eq:4.6}) (``the sinus formula''), \hfill\break
Model II for Eq. (\ref{eq:6.44}) (``the su$_q$(2) irreducible tensor operator 
(ITO) formula''), \hfill\break
Model II$'$ for Eq. (\ref{eq:10.7}) (``the hyperbolic tangent formula''), 
\hfill\break 
Model III for Eq. (\ref{eq:12.1}) (the standand rotational formula), and 
\hfill\break 
Model IV for Eq. (\ref{eq:12.2}) (the Holmberg--Lipas formula).  \hfill\break
It should be emphasized at this point that in models I and I$'$ 
the deformation parameter is a phase factor ($q=e^{i\tau}$, $\tau$ real), 
while in models II and II$'$ the deformation parameter is a real number 
($q=e^\tau$, $\tau$ real). A consequence of this fact is the presence 
of trigonometric functions in models I and I$'$, while in models II and 
II$'$ hyperbolic functions appear. 

The parameters resulting from the relevant least square fits, together with 
the quality measure 
\begin{equation} \label{eq:12.3}
\sigma = \sqrt{ {2\over \ell_{max}} \sum_{i=2}^{\ell_{max}} 
(E_{exp}(\ell)-E_{th}(\ell))^2 },
\end{equation}
where $\ell_{max}$ is the angular momentum of the highest level 
included in the fit, are listed in Table 1, while in Table 2 the 
theoretical predictions of all models for $^{222}$Th are listed together with 
the experimental spectrum. Finally in Tables 3-5 the theoretical predictions 
of models I$'$, II$'$, III, and IV for the rest of the Th isotopes are listed, 
together with the relevant experimental spectra. 

From these tables the following observations can be made: 

a) As seen in Tables 1 and 2, models I and I$'$ give results which are almost 
identical. The same is true for models II and II$'$. We therefore conclude 
that the approximations carried out in Secs. 5 and 10 are very accurate. 
This is the reason that in Tables 3-5 the results of models I and II are 
omitted in favour of models I$'$ and II$'$. 

b) All models give good results for $^{226}$Th-$^{234}$Th, which lie 
in the rotational region, with $R_4$ ratio between 3.136 and 3.308~, 
with model IV giving the best results and model III giving the worst ones,
while in all cases models II and II$'$ are better than models I and I$'$.
It should be noticed, however, that all models tend to underestimate 
the first several levels of the spectra and the last one or two levels, 
while they overestimate the rest of the levels. In other words, all models
``fail in the same way''. 

c) A similar picture holds for the transitional nucleus $^{224}$Th 
($R_4=2.896$) and the near-vibrational nucleus $^{222}$Th ($R_4=2.399$), 
i.e. still model IV gives the best results and model III the worst,
while models II and II$'$ are better than models I and I$'$. However, 
the deviations from the data get much larger, indicating that all these 
models are inappropriate for describing spectra in the vibrational and
transitional regions, in which the presence of a term linear in $\ell$ 
is required, as in the u(5) and o(6) limits of the Interacting Boson 
Model \cite{IBM}.      

These observations lead to the following conclusions: 

a) One can freely use model I$'$ in the place of model I, and model II$'$
in the place of model II, since the relevant approximations turn out to be 
very accurate. Models I$'$ and II$'$ have the advantage of providing 
simple analytic expressions for the energy, the rotational frequency and the 
moment of inertia. 

b) The fact that models II and II$'$ are better than models I and I$'$ 
indicates that within the same symmetry (su$_q$(2) in this case) it is 
possible to construct different rotational Hamiltonians characterized 
by different degrees of agreement with the data. However, these Hamiltonians 
are too ``rigid', in the sense that they can describe only rotational
spectra, while vibrational and transitional spectra are outide their realm. 

Some additional comments on the convergence of the various expansions 
can be made by considering the quantities \cite{BM}
\begin{equation} \label{eq:12.4}
r_1= { {C\over A} \over \left( {B\over A} \right)^2} = {AC\over B^2}, \qquad
r_2 ={ {D\over A} \over \left( {B\over A} \right)^3}= {A^2 D\over B^3}, 
\end{equation}
which refer to the coefficients of the expansion of Eq. (\ref{eq:12.1}). 
Keeping only the first two terms in the Harris formalism for the energy 
and the moment of inertia leads to the values \cite{BM}
\begin{equation} \label{eq:12.5}
r_1^{Harris}= 4, \qquad r_2^{Harris}=24.
\end{equation}
From Eq. (\ref{eq:4.3}) we obtain for model I$'$ 
\begin{equation} \label{eq:12.6}
r_1^{I'}={2\over 5}, \qquad  r_2^{I'} ={3\over 35} , 
\end{equation}
while from Eq. (\ref{eq:10.8}) we obtain for model II$'$
\begin{equation} \label{eq:12.7}
r_1^{II'}={17\over 20}, \qquad r_2^{II'} ={93\over 140}.
\end{equation}
The Taylor expansion of the Holmberg--Lipas formula (Eq. (\ref{eq:12.2})~)
reads
\begin{equation} \label{eq:12.8}
{E\over a}= {b\over 2} \ell(\ell+1) -{b^2\over 8} (\ell(\ell+1))^2
+{b^3\over 16} (\ell(\ell+1))^3 -{5b^4\over 128} (\ell(\ell+1))^4  +\ldots,
\end{equation}
from which one  obtains 
\begin{equation} \label{eq:12.9}
r_1^{IV}=2, \qquad r_2^{IV}=5.  
\end{equation}
We observe that for models I$'$, II$'$, and IV the quality of the fits
is improved as the values of the ratios $r_1$ and $r_2$ get larger. 

Finally, a word of warning: One could think of fitting the experimental 
spectra by Eq. (\ref{eq:12.1}), keeping the first four terms in the expansion, 
and then trying to use the parameter values obtained from fitting several 
nuclei 
in order to determine ``optimal'' values for the ratios $r_1$ and $r_2$ 
from Eq. (\ref{eq:12.4}) as a function of the mass number. This procedure, 
however, is very unsafe, since the values of the parameters $C$ and 
especially $D$ obtained from the fits are very unstable. 

\section{Discussion}

The main results of the present work are the following:  

a) The rotational invariance of the original su$_q$(2) Hamiltonian 
\cite{RRS,PLB251}
under the usual physical angular momentum has been 
proved explicitly and its connections to the formalisms of Amal'sky \cite{Amal}
(``the sinus formula'') and Harris \cite{Harris} have been given. 

b) An irreducible tensor operator (ITO) of rank one under su$_q$(2) has 
been found and used, through $q$-deformed tensor product and $q$-deformed 
Clebsch--Gordan coefficient techniques \cite{STK593,STK690,STK1068,STK1599},  
for the construction of a new Hamiltonian appropriate 
for the description of rotational spectra, the su$_q$(2) ITO Hamiltonian. 
The rotational invariance of this new Hamiltonian under the usual physical 
angular momentum has been proved explicitly. Furthermore, an approximate
simple closed expression (``the hyperbolic tangent formula'') 
for the energy spectrum of this Hamiltonian has been found and its connection 
to the Harris \cite{Harris}  formalism has been demonstrated. 

From the results of the present work it is clear that the su$_q$(2) 
Hamiltonian, as well as the su$_q$(2) ITO Hamiltonian, are complicated 
functions of the Casimir operator of the usual su(2), i.e. of the square
of the usual physical angular momentum. These complicated functions 
possess the su$_q$(2) symmetry, in addition to the usual su(2) symmetry. 
Matrix elements of these functions can be readily calculated in the deformed 
basis, but also in the usual physical basis. A similar study of a
$q$-deformed quadrupole operator is called for. This operator would allow 
the study of multi-band spectra, in analogy to the Elliott model 
\cite{Elliott}, 
as well as the study of BE(2) transition probabilities. Since $q$-deformation 
appears to describe well the stretching effect of rotational nuclear spectra,
ir is interesting to check what its influence on the corresponding B(E2) 
transition probabilities will be. Work in this direction is in progress. 

\section*{Acknowledgements}
The authors acknowledge support from the Bulgarian Ministry of Science 
and Education under Contracts No. $\Phi$-415 and $\Phi$-547.


\newpage 

\begin{table}

\caption{Parameter values and quality measure $\sigma$ (Eq. (\ref{eq:12.3}))
for models I (Eq. (\ref{eq:1.17})), I$'$ (Eq. (\ref{eq:4.6})), II (Eq. 
(\ref{eq:6.44})), II$'$ (Eq. (\ref{eq:10.7})), III (Eq. (\ref{eq:12.1})), 
and IV (Eq. (\ref{eq:12.2})), obtained from least square fits to 
experimental spectra of Th isotopes (shown in Tables 2 and 3).  
Data have been taken from Refs. \cite{Th222} ($^{222}$Th), \cite{Th224}
($^{224}$Th), \cite{Th226} ($^{226}$Th), \cite{Th228} ($^{228}$Th), 
\cite{Cocks} ($^{230}$Th, $^{232}$Th, $^{234}$Th).  
The $R_4=E(4)/E(2)$ ratio for each isotope is also shown. 
}

\bigskip

\centering
\begin{tabular}{l r r r r r r r  }
\hline
\hline 
          & $^{222}$Th & $^{224}$Th & $^{226}$Th & $^{228}$Th & $^{230}$Th &
            $^{232}$Th & $^{234}$Th \\
$R_4$     & 2.399   & 2.896  & 3.136  & 3.235  & 3.271  & 3.283  & 3.308  \\
          &         &        &        &        &        &        &        \\
{\bf Model I}&      &        &        &        &        &        &        \\
$A$ (keV) &  12.577 & 11.855 & 10.047 &  8.873 &  8.149 &  7.437 &  7.845 \\
$10^2\tau$&   4.857 &  5.527 &  4.701 &  4.507 &  3.512 &  3.141 &  3.312 \\
$\sigma$(keV)& 154.213 & 38.135 & 26.404 & 11.601 & 17.074 & 26.700&10.839\\ 
          &         &        &        &        &        &        &        \\
{\bf Model I$'$}&   &        &        &        &        &        &        \\
$A$ (keV) &  12.582 & 11.861 & 10.052 &  8.876 &  8.150 &  7.438 &  7.847 \\
$10^2\tau$&   4.858 &  5.528 &  4.702 &  4.508 &  3.512 &  3.141 &  3.313 \\ 
$\sigma$(keV)& 154.210 & 38.134 & 26.403 & 11.600 & 17.074 & 26.700&10.839\\
          &         &        &        &        &        &        &        \\
{\bf Model II}&     &        &        &        &        &        &        \\
$A$ (keV) &  13.797 & 12.253 & 10.289 &  8.988 &  8.261 &  7.559 &  7.928 \\
$10^2\tau$&   2.156 &  2.229 &  1.858 &  1.728 &  1.351 &  1.218 &  1.260 \\
$\sigma$(keV)& 125.815 & 32.420 & 21.631 &  9.582 & 13.585 & 21.724 &8.204\\
          &         &        &        &        &        &        &        \\ 
{\bf Model II$'$}&  &        &        &        &        &        &        \\
$A$ (keV) &  13.792 & 12.247 & 10.286 &  8.986 &  8.260 &  7.564 &  7.927 \\
$10^2\tau$&   2.155 &  2.228 &  1.858 &  1.727 &  1.351 &  1.220 &  1.260 \\
$\sigma$(keV)& 125.815 & 32.420 & 21.631 &  9.581 & 13.585 & 21.730 &8.204\\
          &         &        &        &        &        &        &        \\ 
{\bf Model III}&    &        &        &        &        &        &        \\
$A$ (keV) &  11.928 & 11.602 &  9.884 &  8.793 &  8.067 &  7.350 &  7.785 \\
$10^2B$(keV)& 0.703 &  0.977 &  0.616 &  0.525 &  0.291 &  0.210 &  0.253 \\ 
$\sigma$(keV)  & 173.357 & 42.528 & 30.137 & 13.238 & 19.912&30.710&13.026\\
          &         &        &        &        &        &        &        \\
{\bf Model IV}&     &        &        &        &        &        &        \\
$10^{-2}a$(keV)&13.812&22.413& 30.636 & 36.853 & 54.344 & 58.577 & 63.701 \\
$10^2b$   &   2.909 &  1.211 &  0.720 &  0.505 &  0.316 &  0.270 &  0.256 \\ 
$\sigma$(keV)  &  53.745 & 18.244 & 10.080 &  4.677 &  5.216 & 9.754&2.139\\
\hline
\end{tabular}
\end{table}

\newpage 

\begin{table}

\caption{Theoretical predictions of models I (Eq. (\ref{eq:1.17})), 
I$'$ (Eq. (\ref{eq:4.6})), II (Eq. (\ref{eq:6.44})), 
II$'$ (Eq. (\ref{eq:10.7})), III (Eq. (\ref{eq:12.1})), 
and IV (Eq. (\ref{eq:12.2})), obtained from least square fits to 
the experimental spectrum (exp.) of $^{232}$Th, taken from Ref. \cite{Cocks}. 
All energies are given in keV. The relevant model parameters and quality 
measure $\sigma$ (Eq. (\ref{eq:12.3})) are given in Table 1. 
}

\bigskip

\centering
\begin{tabular}{ r r r r r r r r  }
\hline
\hline 
       &      &      &       &$^{232}$Th &   &       &    \\
$\ell$ & exp. &  I   & I$'$  & II    & II$'$ & III   &  IV   \\
 2&  49.4&  44.5&   44.5&   45.2&   45.3&   44.0&   47.3\\
 4& 162.2& 147.8&  147.8&  150.0&  150.1&  146.2&  156.3\\
 6& 333.3& 308.1&  308.1&  312.3&  312.5&  305.0&  323.6\\
 8& 557.1& 523.0&  523.0&  529.1&  529.4&  518.3&  544.8\\
10& 826.9& 789.0&  789.0&  796.6&  797.0&  783.0&  814.4\\
12&1136.9&1102.1& 1102.1& 1110.0& 1110.6& 1095.4& 1126.9\\
14&1482.3&1457.1& 1457.1& 1464.2& 1464.8& 1450.8& 1476.7\\
16&1858.3&1848.6& 1848.6& 1853.4& 1854.0& 1843.7& 1858.8\\
18&2261.7&2270.3& 2270.3& 2271.7& 2272.3& 2267.9& 2268.7\\
20&2690.5&2715.7& 2715.7& 2713.1& 2713.6& 2716.4& 2702.3\\
22&3142.9&3177.6& 3177.6& 3171.7& 3171.9& 3181.2& 3156.3\\
24&3618.3&3648.9& 3648.9& 3641.8& 3641.7& 3653.7& 3627.6\\
26&4114.9&4122.0& 4122.0& 4118.0& 4117.9& 4124.5& 4113.9\\
28&4630.5&4589.5& 4589.5& 4595.6& 4594.6& 4583.2& 4613.1\\
\hline
\end{tabular}
\end{table}

\newpage 

\begin{table}

\caption{Theoretical predictions of models I$'$ (Eq. (\ref{eq:4.6})), 
II$'$ (Eq. (\ref{eq:10.7})), III (Eq. (\ref{eq:12.1})), 
and IV (Eq. (\ref{eq:12.2})), obtained from least square fits to 
the experimental spectra (exp.) of $^{222}$Th \cite{Th222} and 
$^{224}$Th \cite{Th224}. 
All energies are given in keV. The relevant model parameters and quality 
measure $\sigma$ (Eq. (\ref{eq:12.3})) are given in Table 1. 
}

\bigskip

\centering
\begin{tabular}{ r r r r r r r r r r r  }
\hline
\hline 
  &      &      &$^{222}$Th&    &       &       &      &$^{224}$Th&  &     \\
$\ell$  & exp. &I$'$  & II$'$ & III   &  IV   & exp.  &  I$'$&II$'$ &III& IV\\
 2& 183.3&  75.1&   82.1&   71.3&  115.7&   98.1&  70.7&  72.9&  69.3&  80.0\\ 
 4& 439.8& 247.7&  269.1&  235.7&  356.0&  284.1& 232.4& 238.6& 228.1& 256.8\\
 6& 750.0& 511.2&  550.4&  488.6&  677.7&  534.7& 477.2& 487.1& 470.1& 511.8\\
 8&1093.5& 855.7&  910.7&  822.3& 1048.6&  833.9& 793.2& 804.1& 784.7& 825.5\\
10&1461.1&1268.3& 1332.0& 1227.0& 1449.6& 1173.8&1164.9&1172.8&1158.0&1181.8\\
12&1850.7&1733.4& 1795.0& 1689.6& 1869.4& 1549.8&1574.4&1575.5&1572.1&1568.8\\
14&2259.7&2233.6& 2281.0& 2194.8& 2301.7& 1958.9&2001.6&1995.4&2005.5&1978.1\\
16&2687.8&2749.9& 2773.1& 2934.3& 2742.5& 2398.0&2425.7&2417.4&2432.8&2403.8\\
18&3133.5&3263.0& 3257.1& 3257.0& 3189.4& 2864.0&2826.2&2829.1&2824.9&2841.7\\
20&3596.0&3753.6& 3721.7& 3769.5& 3640.7&       &      &      &      &      \\
22&4077.6&4203.2& 4158.7& 4235.4& 4095.4&       &      &      &      &      \\
24&4577.9&4594.9& 4562.9& 4625.8& 4552.7&       &      &      &      &      \\
26&5097.9&4914.0& 4931.3& 4908.8& 5012.0&       &      &      &      &      \\
\hline
\end{tabular}
\end{table}

\newpage 

\begin{table}

\caption{Same as Table 3, but for $^{226}$Th \cite{Th226}, 
and $^{228}$Th \cite{Th228}.
}

\bigskip

\centering
\begin{tabular}{ r r r r r r r r r r r  }
\hline
\hline 
  &      &     &$^{226}$Th&     &       &       &      &$^{228}$Th&  &      \\
$\ell$& exp. & I$'$ & II$'$ & III   &  IV   & exp.  & I$'$ &II$'$ &III   &IV\\
 2&  72.2&  60.0&   61.4&   59.1&   65.5&   57.8&  53.0&  53.7&  52.6&  55.4\\ 
 4& 226.4& 198.1&  202.0&  195.2&  213.3&  186.8& 175.1& 176.9& 173.8& 181.7\\
 6& 447.3& 409.3&  415.9&  404.2&  432.9&  378.2& 362.3& 365.1& 360.0& 372.2\\
 8& 721.9& 686.1&  694.2&  679.7&  711.9&  622.5& 608.5& 611.6& 605.9& 618.3\\
10&1040.3&1018.9& 1026.1& 1012.6& 1038.1&  911.8& 905.8& 907.9& 903.7& 911.3\\
12&1395.2&1395.9& 1399.5& 1391.9& 1401.2& 1239.4&1244.4&1244.5&1244.0&1242.6\\
14&1781.5&1803.8& 1802.2& 1803.8& 1792.9& 1599.5&1613.5&1611.3&1615.1&1605.2\\
16&2195.8&2228.1& 2221.9& 2232.5& 2206.9& 1988.1&2001.0&1998.3&2003.4&1993.0\\
18&2635.1&2654.1& 2647.9& 2659.5& 2638.3& 2407.9&2394.5&2396.1&2393.4&2401.2\\
20&3097.1&3066.5& 3070.4& 3064.2& 3083.4&       &      &      &      &      \\
\hline
\end{tabular}
\end{table}

\newpage 

\begin{table}

\caption{Same as Table 3, but for $^{230}$Th \cite{Cocks}, 
and $^{234}$Th \cite{Cocks}. 
}

\bigskip

\centering
\begin{tabular}{ r r r r r r r r r r r  }
\hline
\hline 
  &      &      &$^{230}$Th&    &       &       &      &$^{234}$Th&  &     \\
$\ell$& exp. & I$'$ & II$'$ &  III  &  IV   &  exp. & I$'$ & II$'$& III  &IV\\
 2&  53.2&  48.8&   49.4&   48.3&   51.3&   49.6&  47.0&  47.4&  46.6&  48.7\\ 
 4& 174.0& 161.7&  163.6&  160.2&  169.0&  164.1& 155.8& 157.2& 154.7& 161.1\\
 6& 356.5& 336.4&  340.0&  333.7&  349.3&  337.5& 324.5& 327.1& 322.5& 333.8\\
 8& 593.9& 569.7&  574.5&  565.8&  586.4&  565.7& 550.3& 553.8& 547.4& 562.4\\
10& 879.6& 856.7&  862.1&  852.2&  873.9&  843.5& 829.0& 832.9& 825.7& 841.5\\
12&1207.5&1192.0& 1196.7& 1187.8& 1205.4& 1165.8&1155.8&1159.2&1152.8&1165.6\\
14&1572.8&1568.8& 1571.6& 1565.9& 1574.5& 1527.6&1525.1&1527.0&1523.1&1529.1\\
16&1970.7&1979.8& 1979.7& 1979.2& 1975.6& 1924.4&1930.3&1930.0&1930.1&1926.8\\
18&2397.5&2416.8& 2413.6& 2419.0& 2403.9& 2352.0&2364.3&2361.7&2366.1&2354.1\\
20&2849.8&2871.3& 2866.3& 2875.4& 2855.1& 2806.1&2819.5&2815.6&2822.7&2806.8\\
22&3325.2&3334.3& 3330.8& 3337.7& 3325.7& 3282.4&3288.0&3285.4&3290.5&3281.4\\
24&3820.2&3796.6& 3800.8& 3793.7& 3812.7& 3776.1&3761.5&3765.0&3758.8&3774.8\\
\hline
\end{tabular}
\end{table}

\end{document}